\documentclass[journal]{new-aiaa}
\usepackage[utf8]{inputenc}
\usepackage{textcomp}

\usepackage{graphicx}
\usepackage{amsmath}
\usepackage{bm}
\usepackage{siunitx}
\usepackage{longtable,tabularx}
\usepackage{booktabs}
\usepackage{multirow}
\usepackage{makecell}
\usepackage{subcaption}
\usepackage{algorithm}
\usepackage{algpseudocode}
\usepackage[section]{placeins}
\usepackage{comment}

\hypersetup{colorlinks=true, linkcolor=black, citecolor=black, urlcolor=black}
\makeatletter
\newcommand{\footnoteanchor}{%
  \stepcounter{Hfootnote}%
  \global\let\Hy@saved@currentHref\@currentHref
  \hyper@makecurrent{Hfootnote}%
  \global\let\Hy@footnote@currentHref\@currentHref
  \global\let\@currentHref\Hy@saved@currentHref}
\makeatother

\DeclareSIUnit{\DU}{DU}
\DeclareSIUnit{\TU}{TU}
\DeclareSIUnit{\VU}{VU}
\DeclareSIUnit{\MU}{MU}
\DeclareSIUnit[number-unit-product={}]{\percent}{\%}
\DeclareSIUnit{\flop}{FLOP}

\newcommand{\bs}[1]{\boldsymbol{#1}}
\newcommand{\T}{^\top}
\newcommand{\norm}[1]{\left\lVert #1 \right\rVert}
\newcommand{\snorm}[1]{\lVert #1 \rVert}
\newcommand{\nomsub}{\mathrm{nom}}
\newcommand{\refsub}{\mathrm{ref}}
\newsavebox{\bestvaluebox}
\newsavebox{\bestvalueboldbox}
\newcommand{\bestnum}[1]{%
  \begingroup
  \sbox{\bestvaluebox}{\num{#1}}%
  \sbox{\bestvalueboldbox}{\bfseries\num[detect-weight=true]{#1}}%
  \resizebox*{\wd\bestvaluebox}{\dimexpr\ht\bestvalueboldbox+\dp\bestvalueboldbox\relax}{\usebox{\bestvalueboldbox}}%
  \endgroup
}

\newcounter{appendix}
\renewcommand{\theappendix}{\Alph{appendix}}
\makeatletter
\newcommand{\appendixsection}[1]{%
  \stepcounter{appendix}%
  \section*{Appendix~\theappendix: #1}%
  \phantomsection
  \protected@edef\@currentlabel{\theappendix}}
\makeatother

\title{Neural Costate Guidance for Autonomous Cislunar Low-Thrust Trajectory Generation}

\author{
Jannik Graebner\footnote{Ph.D. Candidate, Department of Mechanical and Aerospace Engineering; jg3607@princeton.edu (Corresponding Author).}
and Ryne Beeson\footnote{Assistant Professor, Department of Mechanical and Aerospace Engineering; ryne@princeton.edu. Senior Member AIAA (Corresponding Author).}
}
\affil{Princeton University, Princeton, NJ 08544}

\begin{document}

\begingroup
\renewcommand{\thefootnote}{}%
\footnoteanchor
\footnotetext{An abbreviated version of this paper has been accepted for presentation at the 37th AAS/AIAA Space Flight Mechanics Meeting, New Orleans, Louisiana, January 2027, as Paper 27-656.}%
\endgroup

\maketitle

\begin{abstract}
Future missions will benefit from onboard autonomous low-thrust trajectory generation to correct deviations from the nominal flight path caused by navigation errors and thruster outages.
Onboard computing resources rarely allow the solution of a cold-started, long-duration nonconvex optimal control problem, especially in dynamically challenging regions such as the cislunar to translunar corridor of the Earth--Moon system.
This paper develops data augmentation and training techniques that enable a neural network to predict accurate initial costates of the indirect method. 
A recovery trajectory is then built from repeated network queries in a computationally efficient guidance algorithm.
Training data are produced by the backward generation of optimal examples method, extending prior approaches to the circular restricted three-body problem.
A novel ``anchor trajectory'' approach is developed to cover the relevant state space, and costate rescaling and a Hamiltonian loss term are used to improve learning.
The method is demonstrated on Earth--Moon transfers from a high Earth orbit to halo orbits about $\mathcal{L}_1$ and $\mathcal{L}_2$.
Under navigation errors and thruster outages, a spacecraft reaches the target orbit more often with the proposed method than with nominal-trajectory tracking or fully numerical reoptimization.
Onboard computation is reduced by more than an order of magnitude compared with reoptimization.
\end{abstract}

\section*{Nomenclature}
\begin{singlespace}
{\renewcommand\arraystretch{1.0}
\noindent\begin{longtable*}{@{}l @{\quad=\quad} l@{}}
$\bs{A}$, $\bs{B}$ & matrices of the linearized dynamics \\
$\bs{a}$ & thrust acceleration, \si{\DU\per\TU\squared} \\
$\Delta\bs{a}$ & feedback acceleration correction, \si{\DU\per\TU\squared} \\
$C$ & Jacobi constant \\
$c$ & exhaust velocity, $I_\mathrm{sp}g_0$, \si{\VU} \\
$D_{\max}$ & maximum initial-state distance from the anchor, canonical units \\
$\bs{d}$ & arrival miss in position and velocity, \si{\DU} and \si{\VU} \\
$d_f$ & arrival error, distance to the target orbit, canonical units \\
$\bs{e}$ & terminal constraint violation, \si{\DU} and \si{\VU} \\
$e_\mathrm{on}$, $e_\mathrm{ramp}$ & start and ramp length of the Hamiltonian penalty, epochs \\
$\bs{f}$ & controlled dynamics \\
$\bs{G}_r$, $\bs{G}_v$ & Jacobians of $\bs{g}$ with respect to $\bs{r}$ and $\bs{v}$, \si{\per\TU\squared} and \si{\per\TU} \\
$\bs{g}$ & natural acceleration, \si{\DU\per\TU\squared} \\
$g_0$ & standard gravity, \si{\meter\per\second\squared} \\
$H$ & Hamiltonian \\
$I_\mathrm{sp}$ & specific impulse, \si{\second} \\
$\bs{I}_3$, $\bs{I}_6$ & $3\times3$ and $6\times6$ identity matrices \\
$J$ & objective \\
$J_\mathrm{LQR}$ & LQR cost \\
$K$ & number of samples per nominal time of flight \\
$k$ & objective scaling parameter \\
$L$ & number of hidden layers \\
$\mathcal{L}$, $\mathcal{L}_H$ & training loss and Hamiltonian penalty \\
$\mathcal{L}_1,\ldots,\mathcal{L}_5$ & libration points \\
$M$ & number of anchor trajectories \\
$m$ & spacecraft mass, \si{\kilo\gram} \\
$m_0$ & initial (wet) mass, \si{\kilo\gram} \\
$m_1$, $m_2$ & primary and secondary masses, \si{\kilo\gram} \\
$N$ & number of trajectories per anchor \\
$\mathcal{N}(\cdot,\cdot)$ & normal distribution \\
$n_r$ & maximum refinement iterations per replan \\
$\bs{P}$, $\bs{P}_f$ & LQR Riccati matrix and its terminal value \\
$\bs{Q}_x$, $\bs{Q}_a$ & LQR state and control weights \\
$R$ & maximum number of replans \\
$\bs{r}$ & position, components $r_1$, $r_2$, $r_3$, \si{\DU} \\
$S$ & switching function \\
$s$ & perturbation multiplier \\
$T$ & thrust magnitude, \si{\newton} \\
$T_\mathrm{max}$ & maximum thrust, \si{\newton} \\
$\mathcal{T}$ & target-orbit period, \si{\TU} \\
$t$ & time, \si{\TU} \\
$t_a$ & anchor epoch, \si{\TU} \\
$t_\mathrm{disp}$ & dispersion epoch, \si{\TU} \\
$t_f$ & final time (time of flight), \si{\TU} \\
$t_{f,\max}$ & maximum shooting time, \si{\TU} \\
$t_\mathrm{go}$ & time to go, \si{\TU} \\
$\Delta t$ & sampling step of the training data, \si{\TU} \\
$\Delta t_c$ & control interval of the tracking baselines, \si{\TU} \\
$\Delta t_r$ & replan interval, \si{\TU} \\
$\Delta t_\mathrm{outage}$, $\Delta t_\mathrm{outage}^{\max}$ & outage duration and its per-segment maximum, \si{\TU} \\
$\mathcal{U}(a,b)$ & uniform distribution on $[a,b]$ \\
$\mathbb{U}$ & set of admissible controls \\
$\bs{u}$ & control, thrust direction and magnitude \\
$\hat{\bs{u}}$ & thrust direction \\
$\bs{v}$ & velocity, components $v_1$, $v_2$, $v_3$, \si{\VU} \\
$v_f$ & magnitude of the terminal velocity, \si{\VU} \\
$W$ & number of units per hidden layer \\
$w_H$ & Hamiltonian penalty weight \\
$\bs{x}$ & state (position, velocity, and mass), \si{\DU}, \si{\VU}, and \si{\kilo\gram} \\
$\Delta\bs{x}$ & deviation from the nominal state, \si{\DU}, \si{\VU}, and \si{\kilo\gram} \\
$\Delta\bs{x}_{rv}$ & position and velocity part of $\Delta\bs{x}$, \si{\DU} and \si{\VU} \\
$\Delta\bs{r}$, $\Delta\bs{v}$ & position and velocity deviations, \si{\DU} and \si{\VU} \\
$\bs{z}$ & shooting decision vector \\
$\alpha$, $\beta$ & in-plane and out-of-plane thrust angles, \si{\radian} \\
$\Gamma$ & target orbit, as a set of states \\
$\gamma$ & relative damping of the refinement \\
$\delta\bs{r}$, $\delta\bs{v}$, $\delta m$ & position, velocity, and mass dispersions, \si{\DU}, \si{\VU}, and \si{\kilo\gram} \\
$\delta r_{\max}$, $\delta v_{\max}$, $\delta m_{\max}$ & dispersion half-widths, \si{\DU}, \si{\VU}, and \si{\kilo\gram} \\
$\delta\bs{\lambda}$, $\delta t$, $\delta\tau_f$ & refinement steps in $\bs{\lambda}$, $t_f$, and $\tau_f$ \\
$\epsilon_a$ & arrival tolerance, canonical units \\
$\epsilon_\mathrm{feas}$ & solver feasibility tolerance, canonical units \\
$\zeta$ & PD damping ratio \\
$\bs{\eta}$ & network target, rescaled costate deviation \\
$\hat{\bs{\eta}}$ & network output \\
$\eta_{\max}$ & maximum dispersion epoch, fraction of $t_{f,\nomsub}$ \\
$\kappa$ & LQR control-weight multiplier \\
$\Lambda$ & dominant eigenvalue of the monodromy matrix \\
$\bs{\lambda}$ & costate, components $\bs{\lambda}_r$, $\bs{\lambda}_v$, $\lambda_m$ \\
$\hat{\bs{\lambda}}$, $\Delta\hat{\bs{\lambda}}$ & predicted costate and costate deviation \\
$\tilde{\bs{\lambda}}$ & costate rescaled to the nominal magnitude \\
$\bs{\lambda}'_{r,f}$ & provisional final position costate \\
$\lambda_v$ & magnitude of the velocity costate \\
$\mu$ & mass ratio \\
$\mu_q$, $\sigma_q$ & mean and standard deviation of network variable $q$ \\
$\nu$ & stability index \\
$\bs{\xi}$ & network input, state deviation and time to go \\
$\rho_1$, $\rho_2$ & distances to the primary and secondary, \si{\DU} \\
$\sigma$ & throttle \\
$\sigma_\tau$, $\sigma_m$, $\sigma_r$, $\sigma_v$ & standard deviations of the terminal perturbations, $\sigma_\tau$ in \si{\TU} and $\sigma_m$ in units of $m_0$ \\
$\bar{\sigma}_\tau$, $\bar{\sigma}_m$, $\bar{\sigma}_r$, $\bar{\sigma}_v$ & their base values \\
$\tau$ & backward time along a training trajectory, \si{\TU} \\
$\tau_a$ & anchor time of flight, \si{\TU} \\
$\tau_f$ & final coast time along the target orbit, \si{\TU} \\
$\bs{\Phi}$ & sensitivity of $\bs{d}$ to the costate \\
$\phi$ & terminal cost \\
$\omega_n$ & PD natural frequency, \si{\per\TU} \\
\multicolumn{2}{@{}l}{Subscripts}\\
$0$ & initial value \\
$\mathrm{disp}$ & dispersed state \\
$f$ & final value \\
$\Gamma$ & target-orbit state \\
$\nomsub$ & nominal trajectory \\
$\refsub$ & reference (nominal or anchor) trajectory \\
$r$, $v$, $m$ & position, velocity, and mass components
\end{longtable*}}
\end{singlespace}

\addtocounter{table}{-1}

\section{Introduction}
The number of planned deep-space and cislunar missions is growing rapidly, and the increasing complexity of these missions places correspondingly higher demands on their guidance, navigation, and control (GNC) systems~\cite{MammarellaEtAl.2025,NesnasEtAl.2021}.
As more missions operate farther from Earth, the traditional ground-in-the-loop approach to GNC operations becomes a bottleneck because ground stations such as the Deep Space Network are increasingly oversubscribed~\cite{ANDRAT2024}.
Unplanned mission opportunities may also be missed for spacecraft not within sight of ground communication or at a distance too far from Earth for transmission delays to be negligible.
Maturing the level of onboard autonomy is therefore increasingly desirable, particularly for operations where communication is delayed, intermittent, or unavailable~\cite{AminiEtAl.2020}.
Simultaneously, a growing number of missions, such as Dawn and Psyche, rely on low-thrust electric propulsion for its fuel efficiency~\cite{Rayman.2006,Oh.2019}.
For these spacecraft, long thrust arcs and low control authority increase both the importance and the difficulty of autonomous trajectory generation.

Traditionally, spacecraft guidance is based on a nominal trajectory, designed on the ground before launch and tracked by the onboard GNC system.
Onboard autonomy has advanced in navigation and trajectory tracking in recent decades.
For example, Deep Space 1 self-determined its orbit from optical images and autonomously planned and executed its correction maneuvers~\cite{Riedel.2000}.
Orion~\cite{GayEtAl.2024} has similarly been designed to use onboard optical navigation and a two-level targeter conditioned on a preflight exhaustive parameter database to execute burns required to return to Earth without ground contact~\cite{MarchandEtAl.2010}.
Their underlying methods, explicit burn guidance laws~\cite{BrandEtAl.1973,GayEtAl.2024} and onboard differential correctors~\cite{MarchandEtAl.2010,ScarrittEtAl.2023,Rayman.2000}, are flight proven and use impulsive burns or small adjustments of the low-thrust profile to return the spacecraft to the nominal target conditions.
For continuous-thrust tracking, another common approach is linear feedback about the nominal.
One such law is the proportional--derivative (PD) control with constant, heuristically tuned gains.
More advanced laws use time-varying gains: in the linear quadratic regulator (LQR) they are derived from a quadratic tracking objective~\cite{Bryson.1975}, and in neighboring optimal guidance from the second variation of the transfer cost about the optimal nominal~\cite{PontaniEtAl.2015a,PontaniEtAl.2015}.
All of these methods enable handling of moderate navigation errors in dynamically quiescent regions by correcting toward a precomputed nominal trajectory.

A sufficiently large deviation, or a moderate one in a dynamically sensitive region, however, either invalidates the linearization about the nominal or forces a return that costs substantially more fuel than flying a newly optimized transfer.
When an unexpected event such as a thruster outage or misfiring leaves the spacecraft in such a state, a new trajectory has to be planned.
Currently, trajectory redesign requires a ground team to generate and validate a new flight path, whose control profile is then uplinked to the spacecraft for execution at a future time. 
This process can take several days and requires maintaining ground communication links. 
Any mission that has flown and demonstrated low levels of GNC autonomy has still required this ground function for trajectory redesign. 
The main factors preventing onboard trajectory redesign are the lack of guarantees on convergence to a feasible flight solution and computational cost.
For long-duration low-thrust transfers in multibody dynamics, the optimal control problem can be highly nonconvex and sensitive to the initial guess.
Iterative solvers may fail to converge, and their convergence time is not bounded a priori~\cite{Betts.1998}.
This makes closed-loop optimization-based approaches, such as model predictive control~\cite{StarekKolmanovsky.2014,ErenEtAl.2017}, hard to reconcile with the hardware and software constraints on board.

Machine learning offers an opportunity to address both barriers: it shifts expensive computation offline and makes available higher-order function approximations in a larger neighborhood about the nominal trajectory~\cite{IzzoEtAl.2019,ShirobokovEtAl.2021,IzzoEtAl.2024}.
In one line of work, the numerical solver is kept and accelerated by learning.
A model trained on previously solved problems supplies initial guesses, so that the solver converges in a fraction of the time.
Such warm starts have proven reliable in two-body transfers, lunar landing, and atmospheric reentry~\cite{Li.2019,Shi.2022}.
Generative models can learn the structure of solutions to more complex transfers in multibody environments and supply initial guesses to single-shooting~\cite{graebner_JAS,Graebner.2026} and multiple-shooting~\cite{TafazzolEtAl.2026} solvers.
However, these models were developed for ground-based mission design, and their computational demands exceed onboard resources.
In a second line of work, the solver is removed and the feedback law itself is learned.
A neural network maps the current state directly to a thrust command~\cite{IzzoEtAl.2024}.
Such policies have been trained by reinforcement learning, for proximity operations under uncertainty~\cite{FedericiEtAl.2021}, robust interplanetary transfers~\cite{doi:10.2514/1.G005794}, and transfers in planar multibody dynamics~\cite{LaFargeEtAl.2021,FedericiEtAl.2023}.
Their training, however, requires substantial reward tuning, and their control does not follow from the necessary conditions for optimality~\cite{FedericiEtAl.2023,HoltEtAl.2026}.
An alternative is behavioral cloning, which trains the policy by supervised regression of the optimal action on trajectories precomputed on the ground~\cite{SanchezSanchez.2018,Tailor.2019,RubinsztejnEtAl.2020}.
Although these policies can be evaluated on board at negligible cost, their low-thrust demonstrations are limited to two-body dynamics and to planar multibody transfers that require little control effort.

In a third line of work, the optimal solution itself is learned with the indirect approach to optimal control. 
Whereas direct methods transcribe the optimal control problem into a parameter optimization problem, indirect methods derive the first-order necessary conditions for optimality and solve the resulting two-point boundary value problem.
A learned model can predict the associated costates, either directly or as the gradient of a learned value function, and the control then follows from the necessary conditions~\cite{doi:10.2514/1.G005254,Singh.2022,EvansEtAl.2025}.
This retains the optimality structure of the problem at the same negligible onboard cost, and it is the approach this work follows.

\begin{figure}
\centering
\includegraphics[width=\linewidth]{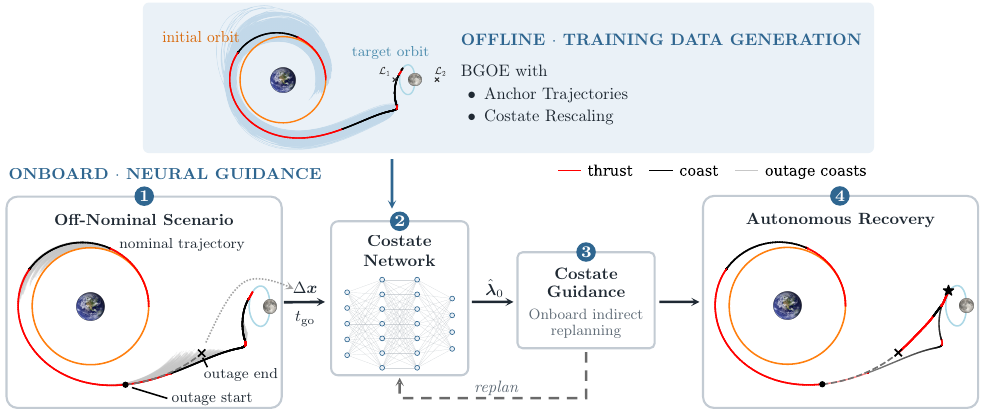}
\caption{Overview of the method on the HEO-to-$\mathcal{L}_1$-halo transfer: offline BGOE and onboard neural guidance with indirect replanning.}
\label{fig:method-overview}
\end{figure}

The indirect formulation offers multiple advantages in a learning context.
First, the solutions satisfy the necessary conditions, so their accuracy is set by the boundary value solver rather than by the resolution (the discretization) of a control transcription.
Second, the unknowns are the initial costates, which have a fixed low-dimensional size independent of the discretization and, most importantly, the time of flight.
Third, the necessary conditions can be exploited to generate training data efficiently.
In particular, solving the boundary value problem once per sample scales poorly, but any trajectory obtained by integrating the state and costate equations backward in time from perturbed terminal conditions satisfying the transversality conditions will also satisfy the other necessary conditions, and therefore be an extremal.
Izzo and Öztürk~\cite{doi:10.2514/1.G005254} named this approach the backward generation of optimal examples (BGOE).
It has since been extended to time-optimal problems~\cite{IzzoOriger.2023}.
Evans et al.~\cite{EvansEtAl.2024} take a different route and approximate the solution of the boundary value problem by polynomial maps obtained through differential algebra, so that guidance updates follow by polynomial evaluation.
In a follow-up study~\cite{EvansEtAl.2025}, the same maps generate the training data for a network that predicts the initial costates and the time of flight.
However, both BGOE and the polynomial maps have so far been formulated and applied only in two-body dynamics.
In this work, we show that the extension to the circular restricted three-body problem (CR3BP) is nontrivial, especially in the cislunar, circumlunar, and translunar corridor. 
It requires modifications to the data generation, the costate representation, and the guidance loop.
Figure~\ref{fig:method-overview} shows an overview of the resulting method: offline, BGOE generates the training data for the costate network, and on board, a guidance loop queries the network to plan the recovery from an off-nominal state.

There are three central contributions of this paper:
\begin{enumerate}
\item 
We extend BGOE to the CR3BP.
Terminal states are perturbed along the target orbit, and the terminal costates are perturbed such that the transversality conditions remain satisfied.
We show that this can be done through an explicit correction of the position costate rather than by root finding, so that each extremal costs only a single backward integration.
\item 
We present two advances in data generation that are essential for guidance in the CR3BP.
Because the three-body dynamics spread the backward-generated extremals unevenly over the region relevant for guidance (i.e., a neighborhood of the phase space enclosing all expected off-nominal states about the nominal trajectory), we generate them not only around the nominal trajectory but also around \emph{anchor trajectories}, which are extremals solved from off-nominal initial states.
It is known that costates are defined only up to a positive factor (positively homogeneous).
We therefore enforce a rescaling of all training costates about anchors to a common magnitude to promote better training and inference. 
\item 
We show that a single network query does not suffice for a reasonably sized network in the dynamically sensitive cislunar region.
We therefore introduce a replanning guidance algorithm, in which the network is repeatedly queried at fixed time intervals and its predictions can be refined by damped differential correction.
\end{enumerate}
Additional contributions of this paper include detailed network hyperparameter and architectural studies and guidance on filtering of data for effective training of the network.

We demonstrate the proposed method on two low-thrust transfers in the Earth--Moon CR3BP from a high Earth orbit (HEO) to halo orbits about $\mathcal{L}_1$ and $\mathcal{L}_2$, both passing through the dynamically sensitive cislunar and circumlunar regions.
The two halo orbits differ in their stability index, and the $\mathcal{L}_2$ transfer is the more challenging of the two, because its target orbit is strongly unstable and lies beyond the Moon, so the transfer must cross the circumlunar region to reach it.
We present a detailed design study, in which we select the relevant data-generation, network, and guidance settings.  
For both transfers we simulate navigation errors and thruster outages. 
We compare the recovery performance of our method with three baseline methods: PD tracking, LQR tracking, and re-solving the full nonconvex optimal control problem numerically.
On the most severe guidance test cases of the $\mathcal{L}_1$ transfer, the spacecraft reaches the target orbit in \SI{94}{\percent} of cases under the network with refinement, compared with \SI{72}{\percent} and \SI{51}{\percent} under PD and LQR tracking of the nominal trajectory.
Our method also requires a factor of \num{60} less onboard computation than re-solving the full nonconvex optimal control problem on board, at comparable fuel cost.
On the second transfer, which targets a strongly unstable halo orbit, the spacecraft still reaches the target orbit more often with our method than with any baseline on the harder test cases, but at lower rates overall.
The main direction for future work is therefore robustness against out-of-distribution queries caused by the growth of small early prediction errors.

The remainder of the paper is organized as follows. 
Section~\ref{sec: background} reviews the CR3BP dynamics, the fuel-optimal transfer problem, and the indirect formulation. 
Section~\ref{sec: problem formulation} defines the two transfers and the guidance test cases.
Section~\ref{sec: methodology} develops the advances in data generation to apply BGOE in the CR3BP and describes the network training and guidance algorithm.
Section~\ref{sec: results} presents the performance results and analysis of the two transfer scenarios, and Sec.~\ref{sec: conclusion} concludes the paper and points to future work.

\section{Background}
\label{sec: background}

\subsection{Low-Thrust Dynamics in the Circular Restricted Three-Body Problem}
\label{sec: low-thrust CR3BP dynamics}
The CR3BP is a widely used idealization for preliminary trajectory design in astrodynamics.
In this model, two massive bodies move in circular orbits about their common barycenter and govern the motion of a third body with negligible mass.
Let the two massive bodies have masses $m_1$ and $m_2$, with $m_1>m_2$, and refer to them as the primary and secondary, respectively.
We adopt the standard system of nondimensional canonical units in which the distance unit (DU) equals the separation between the primaries, the time unit (TU) equals their orbital period divided by $2\pi$, and the mass unit (MU) equals $m_1+m_2$.
The corresponding velocity unit (VU) is \si{\DU\per\TU}.
Under this normalization, the mass ratio $\mu=m_2/(m_1+m_2)$ is the only parameter of the uncontrolled CR3BP.

The equations of motion are expressed in a uniformly rotating reference frame in which the primaries remain fixed on the $r_1$-axis at $r_1=-\mu$ and $r_1=1-\mu$, respectively.
The natural acceleration in this frame includes the gravitational, Coriolis, and centrifugal terms and is given by
\begin{equation}
    \label{eq: CR3BP natural acceleration}
    \ddot{\bs{r}} =
    \begin{pmatrix}
        \ddot{r}_1 \\
        \ddot{r}_2 \\
        \ddot{r}_3
    \end{pmatrix}
    = \bs{g}(\bs{r},\bs{v}) =
    \begin{pmatrix}
        2v_2 + r_1 - (1-\mu)\dfrac{r_1+\mu}{\rho_1^3} - \mu\dfrac{r_1-1+\mu}{\rho_2^3} \\
        -2v_1 + r_2 - (1-\mu)\dfrac{r_2}{\rho_1^3} - \mu\dfrac{r_2}{\rho_2^3} \\
        -(1-\mu)\dfrac{r_3}{\rho_1^3} - \mu\dfrac{r_3}{\rho_2^3}
    \end{pmatrix}
\end{equation}
where $\rho_1=\sqrt{(r_1+\mu)^2+r_2^2+r_3^2}$ and $\rho_2=\sqrt{(r_1-1+\mu)^2+r_2^2+r_3^2}$ denote the distances to the primary and secondary, respectively.
The CR3BP has five equilibrium points, denoted $\mathcal{L}_1$ through $\mathcal{L}_5$, in the rotating frame.
The collinear points $\mathcal{L}_1$ and $\mathcal{L}_2$ lie near the secondary and organize families of three-dimensional periodic halo orbits.
The stability properties of these orbits and the sensitive natural dynamics near the collinear points make accurate guidance particularly important for the transfers considered in this work.

The propulsion system has maximum thrust $T_\mathrm{max}$ and constant specific impulse $I_\mathrm{sp}$, with exhaust velocity $c=I_\mathrm{sp}g_0$, where $g_0$ is the standard gravitational acceleration.
The spacecraft state $\bs{x}=(\bs{r}\T,\bs{v}\T,m)\T\in\mathbb{R}^7$ comprises its position $\bs{r}\in\mathbb{R}^3$, velocity $\bs{v}\in\mathbb{R}^3$, and mass $m$.
With the natural acceleration $\bs{g}(\bs{r},\bs{v})$ of Eq.~\eqref{eq: CR3BP natural acceleration}, the controlled spacecraft dynamics are
\begin{equation}
\label{eq: controlled low-thrust CR3BP dynamics}
    \dot{\bs{x}} = \bs{f}(\bs{x},\bs{u}) =
	\begin{pmatrix}
	\dot{\bs{r}} \\
	\dot{\bs{v}} \\
	\dot{m}
	\end{pmatrix}
	=
	\begin{pmatrix}
	\bs{v} \\
	\bs{g}(\bs{r},\bs{v}) + \dfrac{T}{m}\hat{\bs{u}} \\
	-\dfrac{T}{c}
	\end{pmatrix}
\end{equation}
The control vector consists of the thrust direction and thrust magnitude, $\bs{u}=(\hat{\bs{u}}\T,T)\T$.
The thrust direction is described by the unit vector $\hat{\bs{u}}\in\mathbb{R}^3$, whereas the thrust magnitude is $T=\sigma T_\mathrm{max}$, where the throttle satisfies $\sigma\in[0,1]$.
The set of admissible controls is therefore $\mathbb{U} = \{ (\hat{\bs{u}},T)\in\mathbb{R}^3\times\mathbb{R} \mid \hat{\bs{u}}\T\hat{\bs{u}}=1,\; 0\leq T\leq T_\mathrm{max} \}$.

\subsection{Fuel-Optimal Transfer Problem}
\label{sec: fuel-optimal transfer problem}
We seek an admissible control history $\bs{u}(t)$ and final time $t_f$ that satisfy the controlled dynamics in Eq.~\eqref{eq: controlled low-thrust CR3BP dynamics} and the boundary conditions in Eq.~\eqref{eq: fuel-optimal boundary conditions}.
For a spacecraft with a fixed initial mass, minimizing fuel consumption is equivalent to maximizing the final mass.
We therefore express the minimization objective in Mayer form as
\begin{equation}
    \label{eq: fuel-optimal objective}
    J=\phi(\bs{x}(t_f))=-k m_f
\end{equation}
where $m_f$ denotes the final spacecraft mass and $k>0$ is a fixed scaling parameter.
The trajectory begins from a fixed initial state $\bs{x}_0$ and terminates at a specified position $\bs{r}_f$ and velocity $\bs{v}_f$ on the target orbit.
For each individual transfer problem, the boundary conditions are
\begin{equation}
\label{eq: fuel-optimal boundary conditions}
	\bs{x}(t_0) = \bs{x}_0,
    \qquad
    \bs{e}(\bs{x}(t_f)) =
    \begin{pmatrix}
	\bs{r}_f - \bs{r}(t_f) \\
    \bs{v}_f - \bs{v}(t_f)
    \end{pmatrix} = \bs{0}
\end{equation}
where $\bs{e}$ denotes the terminal constraint violation.
The terminal position and velocity are fixed for each transfer problem, whereas the terminal mass and final time are free.

\subsection{Primer Vector Theory}
\label{sec: primer vector}
Primer vector theory applies Pontryagin's minimum principle to the fuel-optimal transfer problem of the preceding subsection and provides first-order necessary conditions for optimality~\cite{Lawden1964OptimalTF,Liberzon.2012}.
These conditions introduce the costate vector $\bs{\lambda}=(\bs{\lambda}_r\T,\bs{\lambda}_v\T,\lambda_m)\T\in\mathbb{R}^{7}$, consisting of the position, velocity, and mass costates.
Since the objective in Eq.~\eqref{eq: fuel-optimal objective} is in Mayer form, the running cost is zero, and the Hamiltonian is
\begin{equation*}
	H
	=
	\bs{\lambda}\T \bs{f}(\bs{x},\bs{u})
	=
	\bs{\lambda}_r\T \bs{v}
	+
	\bs{\lambda}_v\T
	\left(
	\bs{g}(\bs{r},\bs{v})
	+
	\frac{T}{m}\hat{\bs{u}}
	\right)
	-
	\lambda_m\frac{T}{c}
\end{equation*}
According to Pontryagin's minimum principle~\cite{Pontryagin.2018}, the optimal control minimizes the Hamiltonian over the admissible control set $\mathbb{U}$.
We define the velocity-costate magnitude as $\lambda_v\equiv\snorm{\bs{\lambda}_v}$.
Minimization with respect to the thrust direction yields $\hat{\bs{u}}=-\bs{\lambda}_v/\lambda_v$.
With the switching function $S = \lambda_v + \lambda_m m/c$, the Hamiltonian becomes
\begin{equation}
	\label{eq: Hamiltonian without control}
	H
	=
	\bs{\lambda}_r\T \bs{v}
	+
	\bs{\lambda}_v\T
	\bs{g}(\bs{r},\bs{v})
	-
	S\frac{T}{m}
\end{equation}
To minimize the Hamiltonian over the admissible thrust magnitudes $0\leq T\leq T_\mathrm{max}$, the throttle is chosen on the basis of the sign of $S$, resulting in the discontinuous bang-bang control law~\cite{Prussing.2010}:
\begin{equation}
\label{eq: Control Law}
\hat{\bs{u}}
=
-\frac{\bs{\lambda}_v}{\lambda_v},
\qquad
\sigma
=
\begin{cases}
0 & \text{if } S < 0 \\
1 & \text{if } S > 0 \\
0 \leq \sigma \leq 1 & \text{if } S = 0
\end{cases}
\end{equation}
Applying the first-order necessary conditions for optimality yields the costate dynamics~\cite{Liberzon.2012}
\begin{equation}
\label{eq: Costate equation of motion}
\dot{\bs{\lambda}}
=
\begin{pmatrix}
\dot{\bs{\lambda}}_r \\
\dot{\bs{\lambda}}_v \\
\dot{\lambda}_m
\end{pmatrix}
=
\begin{pmatrix}
-\bs{G}_r\T \bs{\lambda}_v \\
-\bs{\lambda}_r - \bs{G}_v\T \bs{\lambda}_v \\
-\lambda_v T/m^2
\end{pmatrix},
\quad
\text{where}
\quad
\bs{G}_r=\frac{\partial \bs{g}}{\partial \bs{r}},
\qquad
\bs{G}_v=\frac{\partial \bs{g}}{\partial \bs{v}}
\end{equation}
Together with the controlled dynamics in Eq.~\eqref{eq: controlled low-thrust CR3BP dynamics}, the control law in Eq.~\eqref{eq: Control Law}, and the boundary conditions in Eq.~\eqref{eq: fuel-optimal boundary conditions}, the costate dynamics form a two-point boundary value problem.
Since the final time and final mass are free, the corresponding transversality conditions are
\begin{equation}
\label{eq: lam_m=-k}
\lambda_m(t_f)
= \left.\frac{\partial \phi}{\partial m}\right|_{m=m_f}=-k,
\quad \text{and} \quad
H(\bs{x}(t_f), \bs{u}(t_f), \bs{\lambda}(t_f)) = 0
\end{equation}
The CR3BP system is autonomous, and therefore the Hamiltonian is constant along an extremal, so the condition $H(t_f)=0$ implies $H(t)=0$ throughout the trajectory.
The necessary conditions have the property of positive homogeneity. 
Therefore, a positive rescaling of the costates and $k$ does not change the control law.

\section{Problem Formulation}
\label{sec: problem formulation}
This section defines the guidance problem considered in this work.
Two low-thrust transfers in the Earth--Moon CR3BP serve as representative mission cases.
We then specify the off-nominal test cases used to evaluate the neural-network guidance method.

\subsection{Low-Thrust Transfers}
\begin{table}
\caption{Problem-setup parameters for the HEO-to-$\mathcal{L}_1$-halo and HEO-to-$\mathcal{L}_2$-halo transfers. Only the nonzero components of the initial condition are listed.}
\label{tab:problem_setup}
\centering
\setlength{\tabcolsep}{4pt}
\begin{tabular}{llcS[table-format=2.6]S[table-format=2.6]}
\toprule
 & & & \multicolumn{1}{c}{HEO-to-$\mathcal{L}_1$ halo} & \multicolumn{1}{c}{HEO-to-$\mathcal{L}_2$ halo} \\
\midrule
\multirow{2}{*}{\textit{Initial condition}}
& $r_{1,0}$ & \si{\DU} & \multicolumn{2}{c}{\num{0.313028}} \\
& $v_{2,0}$ & \si{\VU} & \multicolumn{2}{c}{\num{1.417856}} \\
\midrule
\multirow{3}{*}{\textit{Target halo orbit}}
& Orbital period $\mathcal{T}$ & \si{\TU} & 2.222 & 3.082 \\
& Jacobi constant $C$ & --- & 2.998 & 3.052 \\
& Stability index $\nu$ & --- & 1.00 & 54.99 \\
\midrule
\multirow{4}{*}{\textit{Spacecraft}}
& Dry mass & \si{\kilo\gram} & \multicolumn{2}{c}{\num{10000}} \\
& Wet (initial) mass $m_0$ & \si{\kilo\gram} & \multicolumn{2}{c}{\num{25000}} \\
& Specific impulse $I_\mathrm{sp}$ & \si{\second} & \multicolumn{2}{c}{\num{3000}} \\
& Max. thrust magnitude $T_\mathrm{max}$ & \si{\newton} & \multicolumn{2}{c}{\num{25.0}} \\
\midrule
\multirow{4}{*}{\textit{Earth--Moon CR3BP}}
& Mass ratio $\mu$ & --- & \multicolumn{2}{c}{\num{0.0121537}} \\
& Distance unit (DU) & \si{\kilo\meter} & \multicolumn{2}{c}{\num{384400}} \\
& Time unit (TU) & \si{\day} & \multicolumn{2}{c}{\num{4.3426}} \\
& Velocity unit (VU) & \si{\kilo\meter\per\second} & \multicolumn{2}{c}{\num{1.0245}} \\
\bottomrule
\end{tabular}
\end{table}
We consider two nominal low-thrust transfers in the Earth--Moon CR3BP: one to a halo orbit about $\mathcal{L}_1$ and one to a halo orbit about $\mathcal{L}_2$.
Both depart from the same near-circular, planar HEO, cross the cislunar domain, and enter the dynamically sensitive circumlunar region organized by the Earth--Moon collinear libration points.
Figure~\ref{fig:outage-trajectories-hard} shows the two nominal trajectories together with some of the outage trajectories defined in Sec.~\ref{sec: guidance test scenarios}.
Each nominal trajectory contains four thrust segments, which are numbered chronologically.

Table~\ref{tab:problem_setup} summarizes the common departure state, target-orbit properties, spacecraft parameters, and Earth--Moon CR3BP constants.
Both transfers use the same spacecraft and propulsion model, whose maximum thrust corresponds to an initial thrust acceleration of \SI{1e-3}{\meter\per\second\squared}.
Unless otherwise noted, trajectory states and times are reported in the canonical distance, velocity, and time units defined in Table~\ref{tab:problem_setup}.
The target halo orbit distinguishes the two mission cases.
The first transfer targets a linearly stable halo orbit about $\mathcal{L}_1$.
The second targets a halo orbit about $\mathcal{L}_2$, beyond the Moon as viewed from Earth.
Although the selected $\mathcal{L}_2$ halo orbit lies at a slightly lower energy level, as reflected by its higher Jacobi constant, it is strongly unstable.
This contrast is quantified by the stability index $\nu = \frac{1}{2}\left( \lvert \Lambda \rvert + \lvert \Lambda \rvert^{-1} \right)$,
where $\Lambda$ denotes the largest-magnitude eigenvalue of the monodromy matrix.
The values $\nu = \num{1.00}$ and $\nu = \num{54.99}$ for the $\mathcal{L}_1$ and $\mathcal{L}_2$ targets, respectively, show one reason why the latter presents a substantially more demanding guidance problem.

\subsection{Guidance Test Cases}
\label{sec: guidance test scenarios}
\begin{figure}[b!]
\centering
\includegraphics[width=\linewidth]{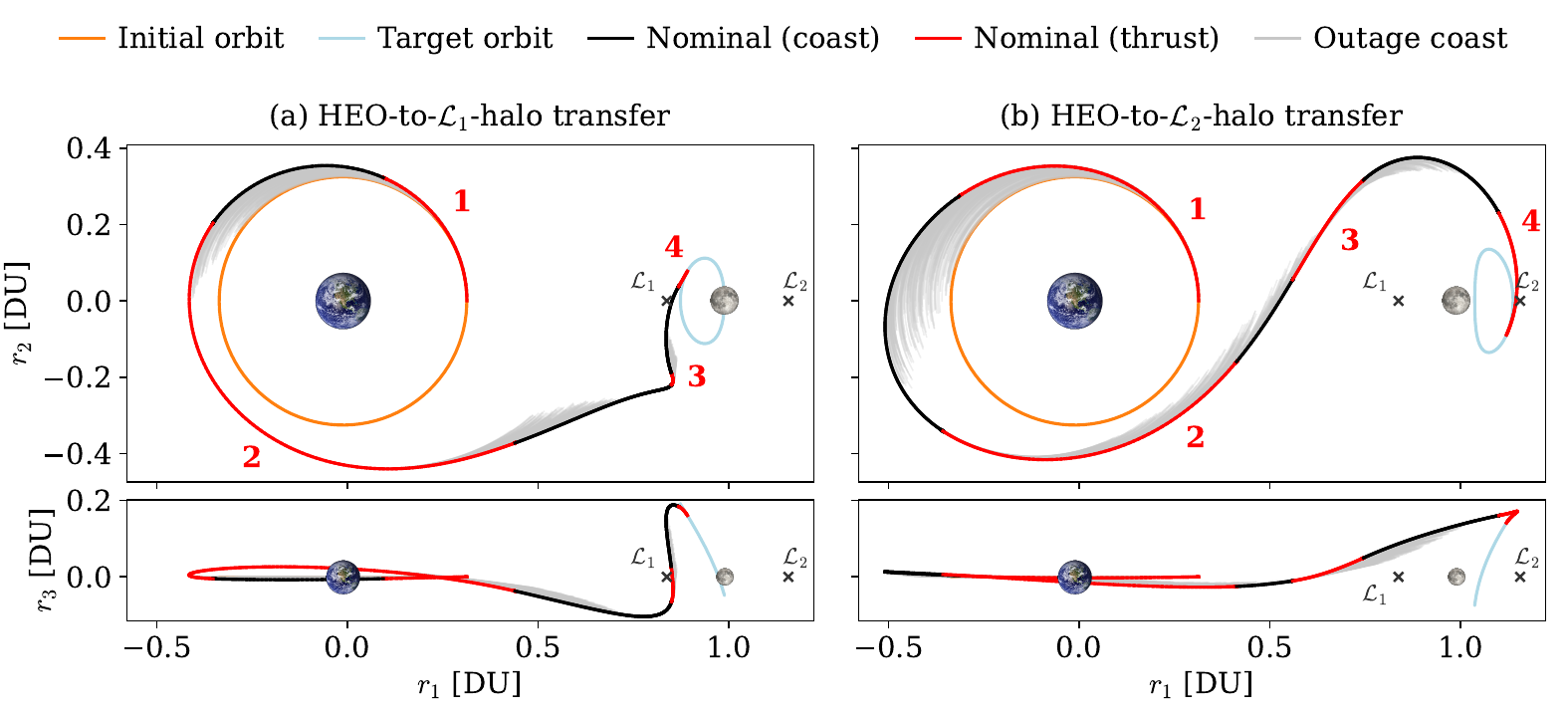}
\caption{Nominal trajectories for both transfers, with \num{3000} outage coasts per transfer from the Hard cases.}
\label{fig:outage-trajectories-hard}
\begingroup
\phantomsubcaption\label{fig:outage-trajectories-heo-l1-halo-hard}
\phantomsubcaption\label{fig:outage-trajectories-heo-l2-halo-hard}
\endgroup
\end{figure}

\begin{table}[t!]
\caption{Guidance test case parameters for the HEO-to-$\mathcal{L}_1$-halo and HEO-to-$\mathcal{L}_2$-halo transfers.}
\label{tab:outage_cases}
\centering
\begin{tabular}{llS[table-format=1.0e-1]S[table-format=1.0e-1]cS[table-format=1.1]S[table-format=1.1]S[table-format=1.1]}
\toprule
& & \multicolumn{3}{c}{Max. initial-state dispersion} & \multicolumn{3}{c}{$\Delta t_\mathrm{outage}^{\max}$ [\si{\TU}]} \\
\cmidrule(lr){3-5} \cmidrule(lr){6-8}
Transfer & Case & \multicolumn{1}{c}{$\delta r_{\max}$ [\si{\DU}]} & \multicolumn{1}{c}{$\delta v_{\max}$ [\si{\VU}]} & $\delta m_{\max}/m_0$ & \multicolumn{1}{c}{Seg.\ 1} & \multicolumn{1}{c}{Seg.\ 2} & \multicolumn{1}{c}{Seg.\ 3} \\
\midrule
\multirow{3}{*}{HEO-to-$\mathcal{L}_1$ halo}
& Easy   & \num{e-3} & \num{e-4} & \num{e-3} & \num{0.0} & \num{0.0} & \num{0.0} \\
& Medium & \num{e-3} & \num{e-4} & \num{e-3} & \num{0.3} & \num{0.4} & \num{0.3} \\
& Hard   & \num{e-3} & \num{e-4} & \num{e-3} & \num{0.6} & \num{0.6} & \num{0.6} \\
\midrule
\multirow{3}{*}{HEO-to-$\mathcal{L}_2$ halo}
& Easy   & \num{e-3} & \num{e-4} & \num{e-3} & \num{0.0} & \num{0.0} & \num{0.0} \\
& Medium & \num{e-3} & \num{e-4} & \num{e-3} & \num{0.4} & \num{0.3} & \num{0.4} \\
& Hard   & \num{e-3} & \num{e-4} & \num{e-3} & \num{0.7} & \num{0.5} & \num{0.7} \\
\bottomrule
\end{tabular}
\end{table}
Our guidance network aims to efficiently generate trajectories that recover from off-nominal states.
We define three test cases of off-nominal states, labeled by difficulty as Easy, Medium, and Hard in Table~\ref{tab:outage_cases}.
Each test case combines two effects:
\begin{enumerate}
    \item Initial-state dispersions are bounded errors in position, velocity, and mass that capture effects such as navigation errors, trajectory insertion errors, maneuver execution errors, or unexpected disturbances.
    \item Thruster outages are thruster malfunctions during a thrust segment of the nominal trajectory, caused, for example, by a propulsion system failure, a loss of power, or a communications interruption.
\end{enumerate}
Each specific test sample is generated by the following procedure:
\begin{enumerate}
\setlength{\emergencystretch}{2em}
\hbadness=1200
    \item Sample a dispersion epoch $t_\mathrm{disp} \sim \mathcal{U}(0,\, \eta_{\max} t_{f,\nomsub})$, where $\mathcal{U}(a,b)$ denotes the uniform distribution on $[a,b]$ and $t_{f,\nomsub}$ is the nominal trajectory's time of flight.
    The factor $\eta_{\max}$ (\num{0.75} for the HEO-to-$\mathcal{L}_1$-halo and \num{0.61} for the HEO-to-$\mathcal{L}_2$-halo transfer) excludes outages during the fourth thrust segment, whose recovery often requires substantially extending the transfer time.
    \item Sample an initial-state dispersion $(\delta\bs{r}\T, \delta\bs{v}\T, \delta m)\T$ with independent Cartesian components $\delta r_i \sim \mathcal{U}(-\delta r_{\max}, \delta r_{\max})$ and $\delta v_i \sim \mathcal{U}(-\delta v_{\max}, \delta v_{\max})$ and a mass component $\delta m/m_0 \sim \mathcal{U}(-\delta m_{\max}/m_0,\, \delta m_{\max}/m_0)$, using the case-specific half-widths of Table~\ref{tab:outage_cases}.
    \item Sample a candidate outage duration $\Delta t_\mathrm{outage} \sim \mathcal{U}(0, \Delta t_\mathrm{outage}^{\max})$, where $\Delta t_\mathrm{outage}^{\max}$ is the maximum outage duration of Table~\ref{tab:outage_cases} for the thrust segment associated with $t_\mathrm{disp}$.
    \item Reject the sample and repeat from step 1 unless the outage interval $[t_\mathrm{disp}, t_\mathrm{disp}+\Delta t_\mathrm{outage}]$ overlaps a thrust segment of the nominal trajectory, because an outage confined to a coast segment leaves the nominal trajectory unaffected.
\end{enumerate}

Given an accepted sample, the dispersion is applied to the nominal state,
\[
\bs{x}_\mathrm{disp} = \bs{x}_\nomsub(t_\mathrm{disp}) + (\delta\bs{r}\T, \delta\bs{v}\T, \delta m)\T
\]
We then propagate $\bs{x}_\mathrm{disp}$ forward under natural, unpowered dynamics for the outage duration $\Delta t_\mathrm{outage}$, holding the mass fixed because the thruster is unavailable over this interval.
The resulting state $\bs{x}_0$ serves as the initial state for guidance replanning.
We assume that full thruster authority is restored instantaneously at the end of the outage and that replanning begins from $\bs{x}_0$ without any additional detection, computation, or actuation delay.

The Easy case includes only the initial-state dispersion and no thruster outage, whereas the Medium and Hard cases include both effects.
For the Medium and Hard HEO-to-$\mathcal{L}_1$-halo cases, we additionally require $t_\mathrm{disp} \geq \SI{1.4}{\TU}$ for outages assigned to segment 2.
We observed empirically that recovery from earlier segment 2 outages often requires additional revolutions about Earth, resulting in a qualitatively different and more complex transfer.
Figure~\ref{fig:outage-trajectories-hard} shows \num{3000} representative Hard-case outage trajectories for each transfer, illustrating the resulting spread of off-nominal states around the nominal trajectory.

\section{Data Generation, Training, and Guidance}
\label{sec: methodology}
This section describes how we generate the training data, train the guidance network, and use the trained network within the guidance algorithm.
We first obtain a nominal trajectory for each mission case through a global search.
We then generate extremals around this trajectory and a set of anchor trajectories through BGOE, and assemble them into a training dataset.
Next, we describe the network architecture and the training procedure.
Finally, we introduce a guidance algorithm that repeatedly queries the trained network to replan along the evolving trajectory.

\subsection{Global Search and Nominal Trajectory Generation}
\label{sec: global search}
The indirect low-thrust trajectory-optimization problem is highly nonconvex and admits numerous locally optimal solutions.
We therefore use a large multistart search to identify high-quality feasible trajectories rather than claim global optimality.
Although the fuel-optimal transfer problem of Sec.~\ref{sec: fuel-optimal transfer problem} minimizes fuel consumption alone, time of flight serves as a secondary criterion for comparing feasible solutions.
Our global-search procedure, summarized briefly here and described fully in our previous work~\cite{graebner_JAS,Graebner.2026}, builds on the work of Russell~\cite{Russell.2007}.
We use indirect forward shooting as implemented in the astrodynamics software \texttt{pydylan}~\cite{Beeson.Aug.2022}.
The sequential quadratic programming solver SNOPT~\cite{Gill.2005} solves the resulting two-point boundary value problem using analytic derivatives.
Each shooting solution is parameterized by the final time and the initial costates, which we collect in the decision vector $\bs{z}=(t_f,\bs{\lambda}_{0}\T)\T$.

Initial guesses for the position and velocity costates are generated using adjoint control transformations (ACT)~\cite{Dixon.1981}.
Evaluating the ACT mapping at the initial time maps $(\alpha_0,\dot{\alpha}_0,\beta_0,\dot{\beta}_0,S_0,\dot{S}_0)$ to $(\bs{\lambda}_{r,0},\bs{\lambda}_{v,0})$, where $\alpha$ and $\beta$ are the in-plane and out-of-plane thrust-vector angles in a velocity-centered spacecraft frame, $S$ is the switching function, and the overdots denote time derivatives at the initial epoch.
The initial mass costate is always initialized at $\lambda_{m,0}=-1$.
We sample these more intuitive initial quantities uniformly instead of sampling the initial costates directly.
The intervals, shared by both transfer cases, are $\alpha_0\in[\num{-0.5},\num{1.1}]\,\si{\radian}$, $\dot{\alpha}_0\in[\num{-9.0},\num{-5.0}]\,\si{\radian\per\TU}$, $\beta_0\in[\num{-1.2},\num{0.3}]\,\si{\radian}$, $\dot{\beta}_0\in[\num{-1.8},\num{2.1}]\,\si{\radian\per\TU}$, $S_0\in[\num{0.0},\num{0.4}]$, and $\dot{S}_0\in[\num{-4.0},\num{0.0}]\,\si{\per\TU}$.
These bounds were refined empirically from broader intervals using the HEO-to-$\mathcal{L}_2$-halo transfer and reused without further tuning for the HEO-to-$\mathcal{L}_1$-halo transfer, where they still yielded a sufficient number of feasible solutions.

A preliminary screening procedure reduces the number of computationally expensive local optimization attempts.
For each initial costate guess, the algorithm propagates the trajectory from the fixed initial state to a maximum shooting time of $t_{f,\max}=\SI{10}{\TU}$ using an adaptive-stepsize RK54 integrator.
At each propagated state, the algorithm identifies the target-orbit state with the nearest Moon-centered phase angle.
A local optimization is attempted only if every component of the corresponding position and velocity differences is smaller than \SI{0.15}{\DU} and \SI{0.15}{\VU}, respectively.
Among the state pairs satisfying these thresholds, the pair with the smallest combined position and velocity mismatch provides the initial guess for $t_f$ and the fixed terminal state.
The resulting problem therefore has a free time of flight and fixed endpoint states but allows different initial guesses to target different locations along the halo orbit.
Further details of this screening procedure are provided in our previous work~\cite{graebner_JAS}.
\begin{figure}
\centering
\begin{subfigure}[b]{0.48\linewidth}
\centering
\includegraphics[width=\linewidth]{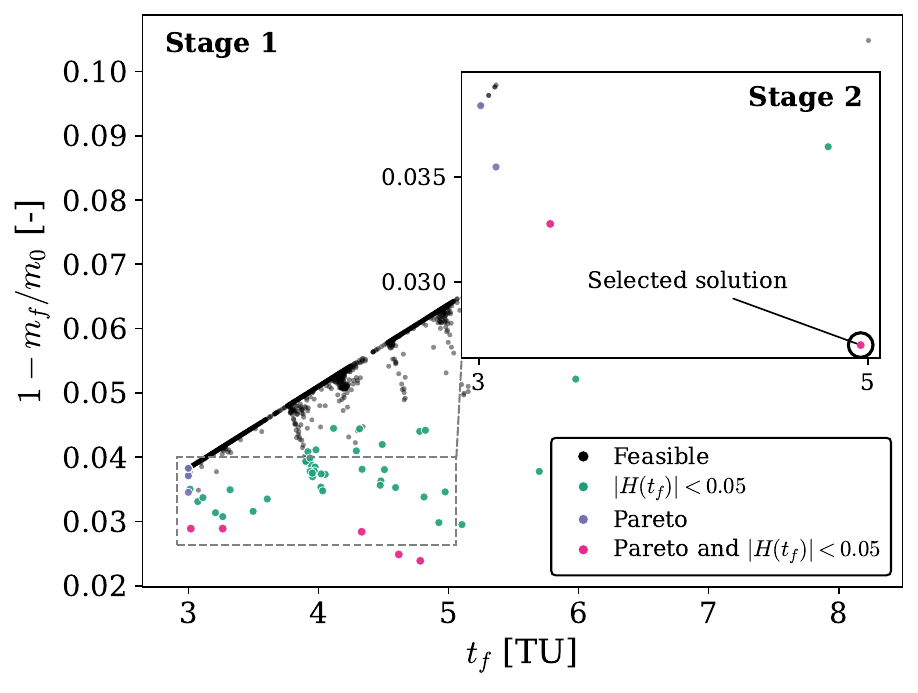}
\caption{HEO-to-$\mathcal{L}_1$-halo transfer}
\label{fig:global-search-heo-l1-halo}
\end{subfigure}
\hfill
\begin{subfigure}[b]{0.48\linewidth}
\centering
\includegraphics[width=\linewidth]{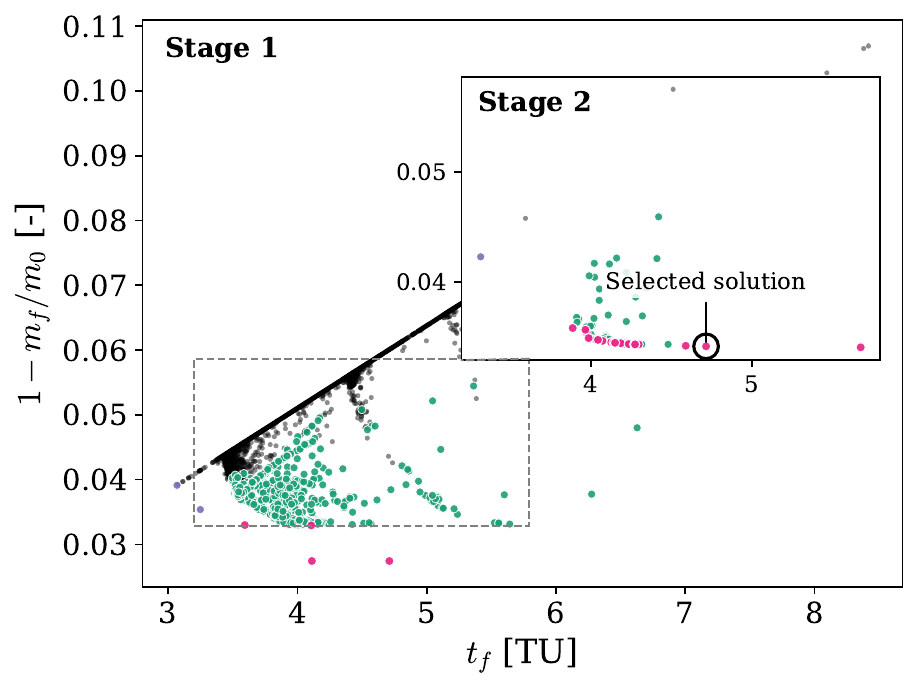}
\caption{HEO-to-$\mathcal{L}_2$-halo transfer}
\label{fig:global-search-heo-l2-halo}
\end{subfigure}
\caption{Two-stage global search in the fuel-consumption and time-of-flight objective space.
The main panel in each subfigure shows the stage 1 feasible-solution cloud, and the inset shows the stage 2 solutions obtained by reoptimizing a subset of stage 1 solutions, with the selected nominal trajectories, shown in Fig.~\ref{fig:outage-trajectories-hard}, circled.
}
\label{fig:global-search}
\end{figure}

We conduct the global search in two stages with progressively tighter SNOPT feasibility tolerances.
For the terminal constraints, a tolerance $\epsilon_\mathrm{feas}$ requires $\snorm{\bs{e}(\bs{x}(t_f))}_\infty\leq\epsilon_\mathrm{feas}$, so every position and velocity residual in canonical units is bounded componentwise.
The feasible solutions from both stages are shown in the time-of-flight and fuel-consumption objective space in Fig.~\ref{fig:global-search} for both transfers.
Stage 1 runs local optimizations from \num{2.8e6} initial guesses with a loose tolerance of $\epsilon_\mathrm{feas}=\num{e-2}$, producing a large set of feasible but only approximately converged solutions.
The $H(t_f)=0$ transversality condition in Eq.~\eqref{eq: lam_m=-k} is not explicitly imposed during this stage.
We retain a stage 1 solution for refinement if it is Pareto-optimal in fuel consumption and time of flight or approximately satisfies the transversality condition, with $\lvert H(t_f)\rvert<\num{0.05}$.
A solution is Pareto-optimal if no other feasible stage 1 solution is at least as good in both objectives and strictly better in at least one.

Stage 2 reoptimizes the retained solutions with a SNOPT feasibility tolerance of $\epsilon_\mathrm{feas}=\num{e-7}$.
During this stage, the Hamiltonian transversality condition is also explicitly enforced with the same tolerance.
The selected HEO-to-$\mathcal{L}_1$-halo trajectory has $(t_f,1-m_f/m_0)=(\SI{4.96}{\TU},\num{0.027})$, whereas the selected HEO-to-$\mathcal{L}_2$-halo trajectory has $(t_f,1-m_f/m_0)=(\SI{4.72}{\TU},\num{0.034})$.
Although this procedure cannot guarantee global optimality, the broad multistart search and tighter second-stage refinement provide high-quality feasible nominal trajectories for the subsequent analysis.

\subsection{Backward Generation of Optimal Examples}
\label{sec: backward generation of optimal examples}
\begin{figure}[b!]
\centering
\includegraphics[width=\linewidth]{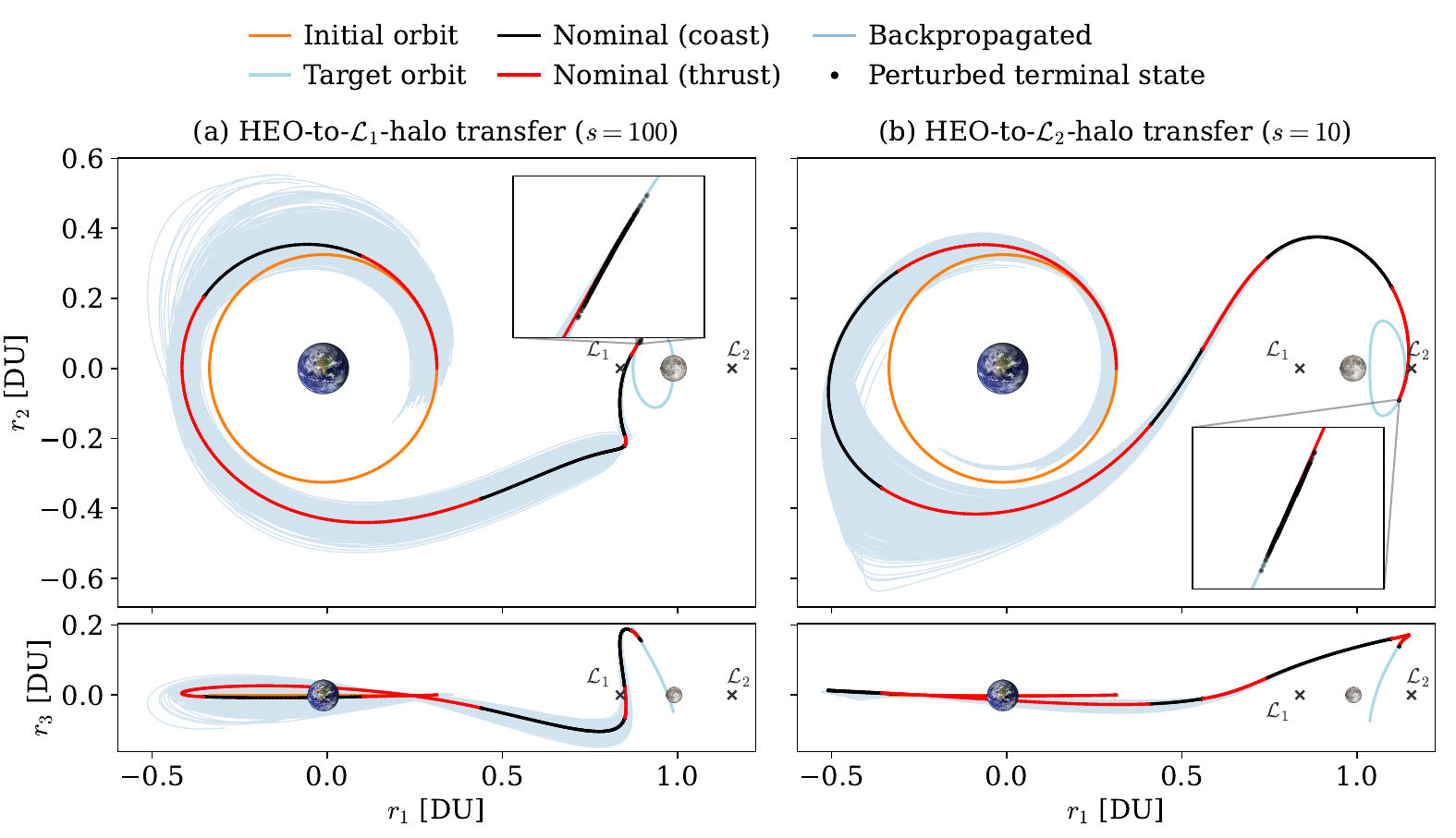}
\caption{Extremals propagated backward over the nominal time of flight from \num{1000} perturbed terminal conditions around each nominal trajectory with perturbation multiplier $s$ as defined in Eq.~\eqref{eq: perturbation multiplier}.}
\label{fig:backprop-trajectories}
\begingroup
\phantomsubcaption\label{fig:backprop-trajectories-heo-l1-halo}
\phantomsubcaption\label{fig:backprop-trajectories-heo-l2-halo}
\endgroup
\end{figure}

Following Izzo and Öztürk~\cite{doi:10.2514/1.G005254}, we generate our dataset by propagating extremals backward in time from perturbed terminal conditions.
Each resulting trajectory satisfies the first-order necessary conditions for optimality by construction, so no optimal control problem has to be solved.
Izzo and Öztürk apply this approach to two-body dynamics in modified equinoctial elements.
We extend it to the CR3BP.
Figure~\ref{fig:backprop-trajectories} illustrates the procedure for both transfers.
Terminal states perturbed along the target orbit (insets) are propagated backward in time, producing a family of extremals around the nominal trajectory.

We describe the procedure for a generic reference trajectory, which is either the nominal trajectory or one of the anchor trajectories introduced in Sec.~\ref{sec: anchor trajectories}.
Quantities of the reference trajectory carry the subscript $\refsub$, and the same symbols without it denote the perturbed trajectory.
We identify each state on the target orbit by its final coast time $\tau_f \in [0, \mathcal{T})$, the time needed to coast along the orbit from that state to a fixed state $\bs{x}_{\Gamma,0}$ on the orbit, where $\mathcal{T}$ is the orbital period.
We denote the state with final coast time $\tau_f$ by $(\bs{r}_\Gamma(\tau_f), \bs{v}_\Gamma(\tau_f))$, so the target orbit is $\Gamma = \{ (\bs{r}_\Gamma(\tau_f), \bs{v}_\Gamma(\tau_f)) \mid \tau_f \in [0, \mathcal{T}) \}$.
Because the reference trajectory is feasible, it ends on $\Gamma$, and we denote its final coast time by $\tau_{f,\refsub}$.
We sample the final coast time, final mass, and final position and velocity costates as
\begin{equation*}
\tau_f \sim \mathcal{N}(\tau_{f,\refsub}, \sigma_\tau^2),
\qquad
m_f \sim \mathcal{N}(m_{f,\refsub}, \sigma_m^2),
\qquad
\bs{\lambda}'_{r,f} \sim \mathcal{N}(\bs{\lambda}_{r,f,\refsub}, \sigma_r^2 \bs{I}_3),
\qquad
\bs{\lambda}_{v,f} \sim \mathcal{N}(\bs{\lambda}_{v,f,\refsub}, \sigma_v^2 \bs{I}_3)
\end{equation*}
and set the perturbed terminal state to $(\bs{r}_f, \bs{v}_f) = (\bs{r}_\Gamma(\tau_f), \bs{v}_\Gamma(\tau_f))$, which lies on the target orbit by construction.
The prime marks the sampled position costate as provisional, because we correct it in Eq.~\eqref{eq:position costate correction} to satisfy the Hamiltonian transversality condition.
Instead of being sampled, the final mass costate is fixed to \mbox{$\lambda_{m,f} = \lambda_{m,f,\refsub} = -k$}, which satisfies the mass transversality condition in Eq.~\eqref{eq: lam_m=-k}.

The sampled costates generally violate the Hamiltonian transversality condition $H(t_f) = 0$ in Eq.~\eqref{eq: lam_m=-k}.
We therefore correct the sampled position costate along $\bs{v}_f$:
\begin{equation}
\label{eq:position costate correction}
\bs{\lambda}_{r,f}
=
\bs{\lambda}'_{r,f}
-
\frac{
H\!\left(
\bs{r}_f,
\bs{v}_f,
m_f,
\bs{\lambda}'_{r,f},
\bs{\lambda}_{v,f},
\lambda_{m,f}
\right)
}{
v_f^2
}
\bs{v}_f
\end{equation}
where $v_f=\snorm{\bs{v}_f}$ denotes the magnitude of the perturbed terminal velocity.
The Hamiltonian in Eq.~\eqref{eq: Hamiltonian without control} is affine in $\bs{\lambda}_r$ through the term $\bs{\lambda}_r\T \bs{v}$, and $\bs{\lambda}_r$ does not enter the switching function or the throttle.
The corrected costate therefore satisfies $H(\bs{x}_f, \bs{\lambda}_f) = 0$ exactly, and the position costate correction in Eq.~\eqref{eq:position costate correction} is the smallest change to $\bs{\lambda}'_{r,f}$ that achieves this.
Izzo and Öztürk require a root-finding step to satisfy this condition in modified equinoctial elements.
In our formulation the correction is explicit, so generating one extremal trajectory costs only a single backward integration.
Propagating $(\bs{x}_f, \bs{\lambda}_f)$ backward in time under the state and costate dynamics and the control law in Eq.~\eqref{eq: Control Law} yields an extremal of the fixed-endpoint, free-final-time problem from its backward-propagated initial state to $\bs{x}_f$.
We integrate backward over the time of flight of the reference trajectory with an adaptive-stepsize RK54 integrator.

\begin{table}[b!]
\caption{Base standard deviations of the terminal perturbations used in BGOE. 
The standard deviations actually used follow from scaling by the multiplier $s$ in Eq.~\eqref{eq: perturbation multiplier}.}
\label{tab:backprop_base_sigmas}
\centering
\begin{tabular}{llS[table-format=1.2e-1]S[table-format=1.2e-1]}
\toprule
Quantity & Unit & \multicolumn{1}{c}{HEO-to-$\mathcal{L}_1$ halo} & \multicolumn{1}{c}{HEO-to-$\mathcal{L}_2$ halo} \\
\midrule
Coast time, $\bar{\sigma}_\tau$ & \si{\TU} & 1.00e-4 & 1.00e-4 \\
Final mass, $\bar{\sigma}_m$ & $m_0$ & 5.93e-5 & 4.00e-5 \\
Position costate, $\bar{\sigma}_r$ & --- & 5.13e-7 & 1.20e-6 \\
Velocity costate, $\bar{\sigma}_v$ & --- & 4.51e-7 & 8.50e-7 \\
\bottomrule
\end{tabular}
\end{table}
Rather than tuning the four standard deviations independently, we tie them to a single multiplier $s>0$:
\begin{equation}
\label{eq: perturbation multiplier}
(\sigma_\tau, \sigma_m, \sigma_r, \sigma_v) = s\,(\bar{\sigma}_\tau, \bar{\sigma}_m, \bar{\sigma}_r, \bar{\sigma}_v)
\end{equation}
The base values $\bar{\sigma}$, listed in Table~\ref{tab:backprop_base_sigmas}, fix the relative size of the four perturbation types for each transfer, and the multiplier $s$ sets the overall perturbation size, which we vary between datasets (Table~\ref{tab:dataset_design}).
We calibrate the base values once per transfer so that each perturbation type, applied alone, causes the same mean deviation of the backward-propagated initial state from the nominal initial state.
For each type, we generate \num{1000} extremals with only that component perturbed and compute the mean of $\sqrt{\snorm{\bs{r}_0-\bs{r}_{0,\nomsub}}^2+\snorm{\bs{v}_0-\bs{v}_{0,\nomsub}}^2}$.
We fix $\bar{\sigma}_\tau$ and iteratively adjust $\bar{\sigma}_m$, $\bar{\sigma}_r$, and $\bar{\sigma}_v$ until their mean deviations match the one caused by $\bar{\sigma}_\tau$.

In the two-body problem considered by Izzo and Öztürk, backward-generated extremals fill an evenly spaced tube around the nominal trajectory and thus cover the nearby states that are relevant for guidance.
In the CR3BP, however, they spread irregularly in different directions, as Fig.~\ref{fig:backprop-trajectories} shows.
Sampling around the nominal trajectory alone therefore leaves large parts of the region that our guidance network should cover empty (Fig.~\ref{fig:anchor-effect-envelopes}).
We instead use BGOE as a data-augmentation method around a small set of additional anchor trajectories, as discussed in the following subsections. 

\subsection{Anchor Trajectories}
\label{sec: anchor trajectories}
We use the anchors to extend the dataset into regions of state space not covered by BGOE around the nominal trajectory alone.
An anchor trajectory is an extremal that connects an off-nominal initial state to the terminal state of the nominal trajectory and satisfies the first-order necessary conditions: the control law in Eq.~\eqref{eq: Control Law}, the costate dynamics in Eq.~\eqref{eq: Costate equation of motion}, and the transversality conditions in Eq.~\eqref{eq: lam_m=-k}.
We then fill the neighborhood of each anchor with BGOE, at the cost of one backward integration per extremal.
The anchor initial states are generated with the sampling procedure of Sec.~\ref{sec: guidance test scenarios}.
We draw them independently of the evaluation test samples, so no test sample serves as the initial state of an anchor.
For both transfers, we sample them from the Hard case, whose off-nominal states contain those of the Easy and Medium cases.

For each anchor initial state $\bs{x}_0$ with associated nominal epoch $t_a$, we solve the fuel-optimal transfer problem of Sec.~\ref{sec: fuel-optimal transfer problem} with the indirect forward-shooting setup of the global search and its stage 2 tolerances, including the explicit $H(t_f)=0$ constraint.
The initial state is fixed to $\bs{x}_0$, the terminal position and velocity are fixed to those of the nominal trajectory, and the time of flight is free but bounded from above by $1.5(t_{f,\nomsub}-t_a)$.
All anchors thus end at the same state on the target orbit, and the perturbation of the final coast time during BGOE later spreads the terminal states of the resulting extremals along the orbit.
We discard solutions that complete an additional revolution about Earth, for the same reason as in Sec.~\ref{sec: guidance test scenarios}.
\begin{figure}[b!]
\centering
\includegraphics[width=\linewidth]{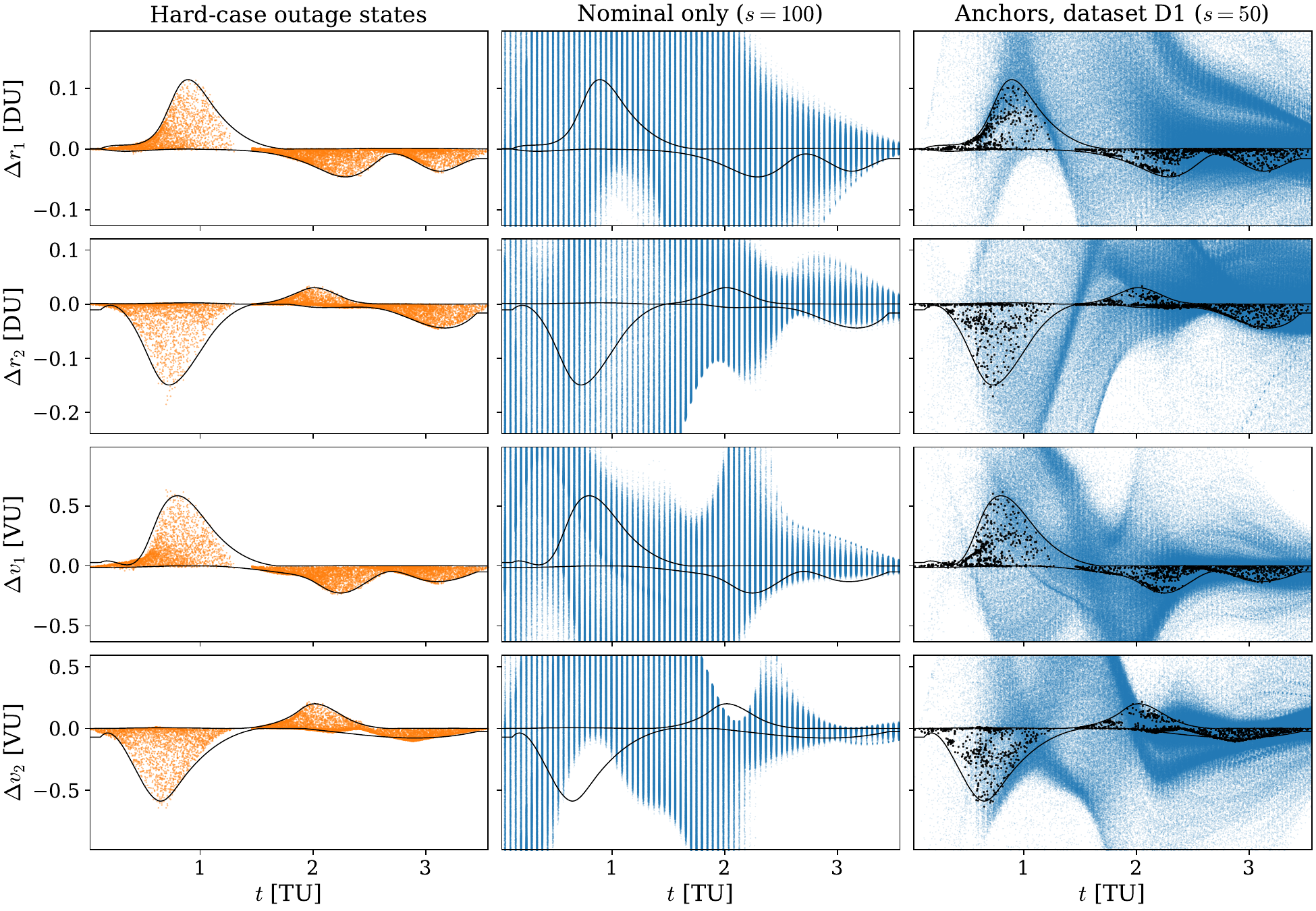}
\caption{Planar position and velocity deviations from the HEO-to-$\mathcal{L}_2$-halo nominal trajectory for the Hard-case outage states and for the datasets generated backward around the nominal trajectory only and around the anchor trajectories (dataset D1 of Table~\ref{tab:dataset_design}).
Each dataset column shows \num{28500} randomly selected trajectories.}
\label{fig:anchor-effect-envelopes}
\end{figure}

We generate the anchors in two stages.
In stage 1, we initialize each problem with the nominal costates at epoch $t_a$ and the remaining nominal time of flight $(t_{f,\nomsub}-t_a)$.
With this warm start and the bound on the time of flight, the anchors remain qualitatively similar to the nominal trajectory.
The converged stage 1 anchors, however, cover the off-nominal states unevenly, because convergence becomes less likely as $\bs{x}_0$ moves farther from the nominal trajectory.
In stage 2, we fill the sparsely covered regions.
We select additional initial states in these regions and initialize each problem with the nearest stage 1 anchor's solution instead of the nominal trajectory.
The nearest anchor is already close to the new solution, so this warm start converges in many cases where the nominal warm start fails.
From the converged anchors of both stages, we select the same number for each thrust segment in which an outage can occur, yielding $M$ anchors per transfer.
Generating $M=\num{3000}$ anchors takes about \num{2000}~CPU hours for the HEO-to-$\mathcal{L}_1$-halo transfer and accounts for most of the cost of dataset generation.
Generating $N=\num{100}$ BGOE trajectories around each of these anchors adds only about \num{2}~CPU hours, which is negligible in comparison.
This difference highlights the importance of BGOE.
Without it, generating a dataset large enough to exploit the advantages of deep learning would be computationally impractical.

Figure~\ref{fig:anchor-effect-envelopes} shows the effect of the anchors on the dataset distribution for the HEO-to-$\mathcal{L}_2$-halo transfer.
Each panel shows the sample deviations from the nominal trajectory in one planar position or velocity component.
The out-of-plane components and the mass are omitted for brevity.
The left column shows the off-nominal states of the Hard case together with an envelope that bounds them at each time.
We obtain this envelope, drawn in black, by fitting smoothing splines to the time-local \SI{0.1}{\percent} and \SI{99.9}{\percent} quantiles of the states, and we plot it in all three columns for comparison.
A training dataset should fill the envelope completely.
Otherwise, the guidance network is queried outside its training distribution.
The dataset generated backward around the nominal trajectory alone (center, $s=\num{100}$) leaves parts of the envelope empty, most visibly in $\Delta v_2$, whereas the anchor dataset (right) covers them.
In the right column, the black points mark only the initial states of the \num{1500} anchors, and the blue points are samples of the perturbed trajectories generated around them through BGOE.
Although the envelope coverage of the anchors alone is still not perfectly even, these perturbed trajectories spread around each anchor and fill the space between them.
A larger perturbation scale would fill more of the gaps in the center column, but it would add many states far from the nominal trajectory, which have been found to degrade training.

This visual comparison does not fully characterize dataset quality for two reasons.
First, the samples lie in an eight-dimensional space of state and time, so coverage of each two-dimensional projection onto one state component and time is necessary but not sufficient for coverage of the full distribution.
A measurement in the full space confirms the advantage of the anchor dataset.
With equally many trajectories drawn from each dataset, adding the anchors reduces the mean normalized distance from a Hard-case outage state to the nearest dataset sample at the same epoch by a factor of \num{2.2} for the HEO-to-$\mathcal{L}_2$-halo transfer and \num{5.6} for the HEO-to-$\mathcal{L}_1$-halo transfer.
Second, the outage states are only the initial states at which the network is first queried.
The algorithm of Sec.~\ref{sec: guidance algorithm} queries the network again along the evolving trajectory, at states that mostly lie outside the envelope.
The samples outside the envelope in the right column are, therefore, important for performance on these later queries.

\subsection{Costate Training Dataset}
\label{sec: costate training dataset}
This subsection describes how we assemble the training dataset from the backward-generated extremals around the anchor trajectories.
Each trajectory yields many training samples, because the remaining trajectory from every state--costate pair along it is an extremal of the fuel-optimal transfer problem started at that state.
We sample all trajectories uniformly in time with the common step $\Delta t=t_{f,\nomsub}/K$, where $K=\num{100}$.
The backward propagation returns the state only at discrete times, so we take the stored state closest to each sample time rather than interpolating between stored states.
More elaborate sampling schemes are possible, but we did not find them necessary for these transfers.
A trajectory perturbed from an anchor with time of flight $\tau_a$ is propagated backward over that same duration and therefore contributes about $K\,\tau_a/t_{f,\nomsub}$ samples.
Each of the $M$ anchors carries $N$ perturbed terminal boundary conditions, so the dataset holds $MN$ trajectories, and Table~\ref{tab:dataset_design} reports the resulting sample count after the filtering steps that follow.

Some samples are either nonphysical or ambiguous in the sense that nearby states carry markedly different costates, and we found that removing them improves training performance.
Four filters remove such samples:
\begin{enumerate}
    \item The first filter removes the last three samples of each backward-generated trajectory.
    The guidance algorithm never queries the network that close to the target orbit.
    These samples also degrade training, because the perturbed costates have not yet spread the trajectories apart.
    \item The second filter removes the samples on a terminal coast arc, together with the three thrusting samples immediately preceding it.
    On a terminal coast the control is zero and no later thrust arc constrains the costate, so different costates can lead to the same trajectory.
    \item The third filter removes complete trajectories whose Hamiltonian drifts to an absolute value greater than $\num{5e-4}$.
    The Hamiltonian is constant along an extremal, so such a drift marks a nonphysical trajectory produced by numerical error, most often in sensitive regions such as a close lunar passage.
    \item The fourth filter removes complete trajectories whose backward-propagated initial state lies at a position--velocity distance greater than $D_{\max}=\num{1.5}$ from the initial state of their anchor trajectory.
    Such trajectories are not relevant for guidance, and this filter acts mostly on the more sensitive HEO-to-$\mathcal{L}_2$-halo case.
\end{enumerate}
Together these filters remove \SI{5.37}{\percent} of the initial samples from the HEO-to-$\mathcal{L}_1$-halo dataset D5 defined in Table~\ref{tab:dataset_design} (\SI{5.33}{\percent} last samples, \SI{0}{\percent} terminal coast, \SI{0}{\percent} Hamiltonian drift, and \SI{0.04}{\percent} divergence) and \SI{35.67}{\percent} from the corresponding HEO-to-$\mathcal{L}_2$-halo dataset (\SI{3.56}{\percent}, \SI{0.58}{\percent}, \SI{0.18}{\percent}, and \SI{31.35}{\percent} in the same order).

Six training datasets with varying parameters are generated through this procedure and summarized in Table~\ref{tab:dataset_design}.
Datasets D1 to D5 cover both transfers, and D6 covers the HEO-to-$\mathcal{L}_1$-halo transfer only.
Section~\ref{sec: design studies} compares them.
The generated datasets use $M=\num{3000}$ anchors for the HEO-to-$\mathcal{L}_1$-halo transfer and $M=\num{1500}$ anchors for the HEO-to-$\mathcal{L}_2$-halo transfer.
A design study (Appendix~\ref{sec: appendix design studies}) indicates that further increasing $N$ or $M$ does not improve performance on the HEO-to-$\mathcal{L}_2$-halo transfer, which is the reason for the smaller $M$ and the omission of D6. 
The datasets differ in the number of perturbed trajectories per anchor $N$ and in the perturbation multiplier $s$, which scales the transfer-specific base standard deviations of Table~\ref{tab:backprop_base_sigmas}.
Dataset D1 uses one value of $s$ for the three thrust segments in which an outage can occur, whereas D2 to D6 use a different $s$ for each segment.
The deviation from the anchor trajectory grows the further back in time we propagate, so perturbations around anchors in segment 1 need a smaller $s$ than those around anchors in segment 3 to reach a similar distance spread.
The HEO-to-$\mathcal{L}_2$-halo datasets are smaller because they use half as many anchors and the filters remove a larger fraction of their samples.
For both transfers the filters remove more samples as $s$ increases.

\begin{table}[b!]
\caption{Design and size of the training datasets.
Each dataset uses $N$ perturbed trajectories per anchor with perturbation multiplier $s$ of Eq.~\eqref{eq: perturbation multiplier} on the $M$ anchors of the transfer.
Dataset size counts the state--costate samples that remain after filtering.}
\label{tab:dataset_design}
\centering
\begin{tabular}{cS[table-format=3.0]cccS[table-format=2.2]S[table-format=2.2]}
\toprule
 & & \multicolumn{3}{c}{$s$} & \multicolumn{2}{c}{Dataset size [$10^6$]} \\
\cmidrule(lr){3-5}\cmidrule(lr){6-7}
Dataset & \multicolumn{1}{c}{$N$} & Seg.~1 & Seg.~2 & Seg.~3
  & {HEO-to-$\mathcal{L}_1$ halo} & {HEO-to-$\mathcal{L}_2$ halo} \\
 & & & & & {($M=\num{3000}$)} & {($M=\num{1500}$)} \\
\midrule
D1 & 100 & \num{50} & \num{50} & \num{50} & 15.95 & 5.45 \\
D2 & 100 & \num{10} & \num{50} & \num{100} & 15.96 & 6.17 \\
D3 & 100 & \num{20} & \num{100} & \num{200} & 15.95 & 5.81 \\
D4 & 200 & \num{10} & \num{50} & \num{100} & 31.92 & 12.34 \\
D5 = D2 $\cup$ D3 & 200 & \num{10}, \num{20} & \num{50}, \num{100} & \num{100}, \num{200} & 31.91 & 11.97 \\
D6 & 400 & \num{10}, \num{20} & \num{50}, \num{100} & \num{100}, \num{200} & 63.82 & \multicolumn{1}{c}{---} \\
\bottomrule
\end{tabular}
\end{table}

After filtering, we prepare the dataset in three steps.
The first step expresses each sample relative to the nominal trajectory.
Every sample carries the time $\tau\in[0,\tau_a]$ that the backward propagation has integrated to reach it, starting from the perturbed terminal state on the target orbit.
Equivalently, $\tau$ is the time to go along the anchor trajectory.
It is zero at arrival on the target orbit and grows to $\tau_a$ at the furthest-back point, so it runs backward in physical time.
The anchor's total time to go $\tau_a$ differs from the nominal time to go $t_{f,\nomsub}-t_a$ at the anchor epoch $t_a$.
We therefore rescale it linearly onto the nominal interval, $t_\mathrm{go}(\tau)=\tau\,(t_{f,\nomsub}-t_a)/\tau_a$, so that the two trajectories are compared at the same fraction of their remaining flight rather than at equal elapsed times.
The corresponding nominal epoch is $t=t_{f,\nomsub}-t_\mathrm{go}$, at which we obtain the nominal state $\bs{x}_{\nomsub}(t)$ and costate $\bs{\lambda}_{\nomsub}(t)$ by linear interpolation of the saved nominal history.
The network input is the state deviation $\Delta\bs{x}=\bs{x}(t)-\bs{x}_{\nomsub}(t)$ together with the time to go, so that $\bs{\xi}=(\Delta\bs{x}\T,\,t_\mathrm{go})\T\in\mathbb{R}^{8}$.
The time entry encodes which part of the transfer a deviation belongs to.

The second step fixes the scaling of the costates.
As stated in Sec.~\ref{sec: primer vector}, the control depends on the costate only through the thrust direction $-\bs{\lambda}_v/\lambda_v$ and the sign of the switching function $S$.
The costate dynamics in Eq.~\eqref{eq: Costate equation of motion} are positively homogeneous of degree one in $\bs{\lambda}$, so scaling $\bs{\lambda}$ by any positive factor yields another extremal with the same control and the same state trajectory.
Solving for the anchors leaves this scale free, so two samples at nearly the same state can carry costates that differ by a large positive factor.
The state-to-costate map is then multivalued, and a network trained under a mean-squared-error (MSE) loss fits the average of the two costates rather than either branch.
We remove this freedom by \emph{costate rescaling}.
Before the nominal costate is subtracted, each training costate is scaled to the magnitude of the nominal costate at the same time, keeping its direction,
\begin{equation}
	\label{eq: costate rescaling}
	\tilde{\bs{\lambda}}(t)=\frac{\norm{\bs{\lambda}_{\nomsub}(t)}}{\norm{\bs{\lambda}(t)}}\,\bs{\lambda}(t)
\end{equation}
Positive multiples of the same costate now map to one $\tilde{\bs{\lambda}}$, which resolves the multivaluedness.
The nominal costate is a fixed point of Eq.~\eqref{eq: costate rescaling}, so costate rescaling leaves the nominal trajectory unchanged and is compatible with the network target $\bs{\eta}=\tilde{\bs{\lambda}}(t)-\bs{\lambda}_{\nomsub}(t)\in\mathbb{R}^{7}$.
Skipping this costate rescaling step leads to significantly worse guidance performance on both transfers, as shown by the ablation studies in Sec.~\ref{sec: design studies} and Appendix~\ref{sec: appendix design studies}.
Cheng et al.~\cite{ChengEtAl.2019} encountered the same ambiguity when training networks on normalized costates and removed it by dividing every costate by its normalization factor.
Costate rescaling removes only one degree of freedom, which is why the terminal-coast samples are filtered out rather than rescaled.
There, any costate whose switching function remains nonpositive throughout the coast produces the same trajectory, so the ambiguity spans a whole region of costate space rather than a single scaling direction.

The third step standardizes every component $q$ of $\bs{\xi}$ and $\bs{\eta}$, replacing it by $(q-\mu_q)/\max(\sigma_q,\num{e-8})$, where $\mu_q$ and $\sigma_q$ are the componentwise mean and standard deviation of the dataset and the lower bound \num{e-8} guards against a vanishing spread.
We compute these statistics on the training split alone, so that no information from the validation or test trajectories enters the transformation.

\subsection{Network Architecture and Training}
\label{sec: network architecture and training}
\begin{figure}[b!]
\centering
\includegraphics[width=\linewidth]{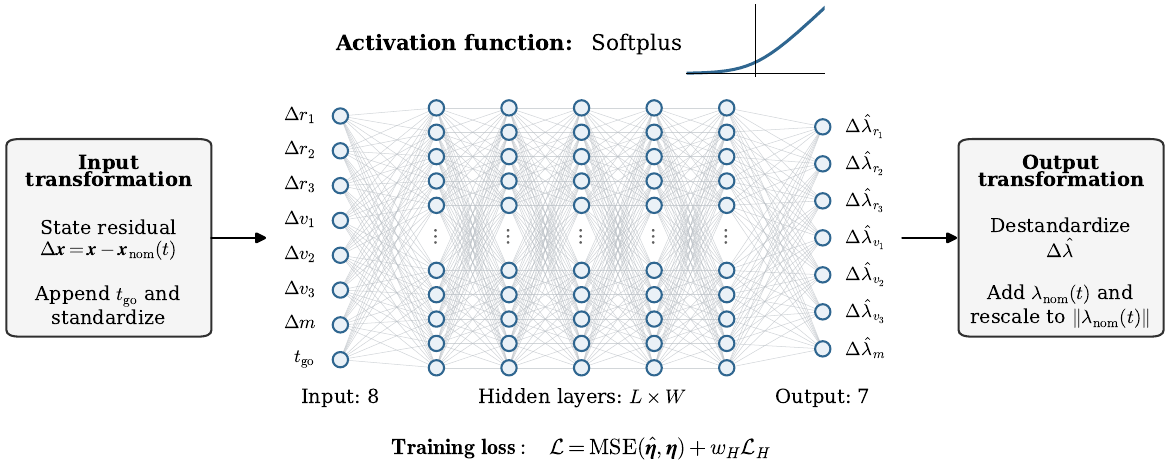}
\caption{Costate-network architecture and transformations.}
\label{fig:costate-network-architecture}
\end{figure}

The guidance network is a fully connected feedforward network that maps the eight standardized inputs $\bs{\xi}$ of Sec.~\ref{sec: costate training dataset} to the seven standardized costate residuals $\bs{\eta}$.
Initial architecture and hyperparameter choices follow Izzo and Öztürk~\cite{doi:10.2514/1.G005254}, with some adaptations to our problem.
The network has $L$ hidden layers of $W$ units each, a Softplus activation on every hidden layer, and a linear output layer.
We train networks with values of $W\in\{\num{100},\num{200},\num{400}\}$ and $L\in\{3,5,7,9\}$, which are compared in Sec.~\ref{sec: model size comparison}.
The weights are initialized with the Kaiming normal scheme~\cite{He.2015}, and all biases are set to zero.
For a prediction, we standardize the input $\bs{\xi}$ with the transformations of Sec.~\ref{sec: costate training dataset}, evaluate the network, and invert the transformations to destandardize its output to a costate deviation $\Delta\hat{\bs{\lambda}}$.
Adding the nominal costate and reapplying the costate rescaling of Eq.~\eqref{eq: costate rescaling} gives the predicted costate $\hat{\bs{\lambda}}=\snorm{\bs{\lambda}_{\nomsub}}(\Delta\hat{\bs{\lambda}}+\bs{\lambda}_{\nomsub})/\snorm{\Delta\hat{\bs{\lambda}}+\bs{\lambda}_{\nomsub}}$.
Figure~\ref{fig:costate-network-architecture} summarizes the network and the transformations around it.

The training objective adds a penalty to a plain regression loss.
Every extremal in the dataset satisfies the condition $H=0$ of Eq.~\eqref{eq: lam_m=-k}, and the penalty holds the prediction to it.
The Hamiltonian is positively homogeneous of degree one in the costate, so we penalize the scale-invariant residual $H/\snorm{\hat{\bs{\lambda}}}$, which cannot be reduced by shrinking the prediction instead of correcting its direction.
The objective is then
\begin{equation}
	\label{eq: training loss}
	\mathcal{L}
	=
	\operatorname{MSE}\left(\hat{\bs{\eta}},\bs{\eta}\right)
	+
	w_{H}\,\mathcal{L}_{H}
\end{equation}
where $\hat{\bs{\eta}}$ is the network output, $\bs{\eta}$ is the stored target, and $\mathcal{L}_{H}$ is the minibatch mean of $(H(\bs{x},\hat{\bs{\lambda}})/\snorm{\hat{\bs{\lambda}}})^{2}$, with $\hat{\bs{\lambda}}$ the costate that the output transform reconstructs from $\hat{\bs{\eta}}$, and the MSE is averaged over the same minibatch.
The penalty needs no data beyond the state already in the sample, and it nudges network predictions toward the manifold on which the data lie.
We ramp $w_H$ linearly from zero to its final value, starting at epoch $e_\mathrm{on}$ and reaching full weight $e_\mathrm{ramp}$ epochs later.
For the HEO-to-$\mathcal{L}_1$-halo transfer, we hold the penalty off for the first $e_\mathrm{on}=\num{50}$ epochs and then ramp it over $e_\mathrm{ramp}=\num{20}$ epochs to $w_H=\num{1000}$, whereas for the more sensitive HEO-to-$\mathcal{L}_2$-halo transfer, we engage it immediately, $e_\mathrm{on}=\num{1}$, but ramp it over $e_\mathrm{ramp}=\num{150}$ epochs to a much smaller $w_H=\num{0.01}$.
The gap between the two weights reflects the scale of the Hamiltonian residual, not a different emphasis on the constraint.
The residual is roughly \num{500} times larger for the HEO-to-$\mathcal{L}_2$-halo transfer, so the smaller $w_H$ leaves the penalty a comparable share of the objective.
These settings are what worked empirically for each transfer and are not necessarily optimal.

We split the dataset into \SI{85}{\percent} training, \SI{10}{\percent} validation, and \SI{5}{\percent} test.
We split at the level of whole anchors, assigning each anchor with all its trajectories and samples to exactly one split.
Samples along a backward-generated trajectory are strongly correlated, and so are the trajectories from one anchor, so a sample-level split would leave near-duplicates of training samples in the validation and test sets.
The split is stratified by thrust segment, so that every set holds the same share of each segment's anchors.

We train each network for \num{250} epochs with the Adam optimizer~\cite{Kingma.2015} in its AMSGrad variant~\cite{Reddi.2018}, using a minibatch size of \num{4096}, an initial learning rate of \num{e-4}, the default moment coefficients, and no weight decay.
Whenever 10 consecutive epochs pass without the validation regression loss improving by more than \SI{0.5}{\percent}, we multiply the learning rate by \num{0.7}.
Training the final HEO-to-$\mathcal{L}_1$-halo network of Sec.~\ref{sec: final guidance performance} ($W=\num{400}$, $L=\num{7}$, dataset D6) takes about \SI{4.1}{\hour} on one NVIDIA L40 GPU.
We keep the epoch with the lowest validation regression loss, and we always train three networks with different random seeds, so the design studies report the spread over these seeds.

\subsection{Guidance Algorithm}
\label{sec: guidance algorithm}
We employ a guidance algorithm that queries the guidance network repeatedly and replans from the propagated state at fixed intervals.
The whole loop runs on board in simulation before the spacecraft flies the trajectory that the loop returns, so we use it as a trajectory-generation tool with a limited number of replans rather than as closed-loop guidance.
The trained network could also be queried in flight from the navigated state as closed-loop guidance, but this is not the focus of this work.
This subsection motivates and then describes the algorithm.

To evaluate the performance of the algorithm with the guidance network, we measure how closely a trajectory reaches the target orbit by its arrival error
\begin{equation}
d_f=\min_{\tau_f\in[0,\mathcal{T})}\norm{\begin{pmatrix}\bs{r}(t_f)-\bs{r}_\Gamma(\tau_f)\\ \bs{v}(t_f)-\bs{v}_\Gamma(\tau_f)\end{pmatrix}}
    \label{eq: arrival error}
\end{equation}
where $\bs{r}_\Gamma(\tau_f)$ and $\bs{v}_\Gamma(\tau_f)$ are the position and velocity of the target-orbit state and $t_f$ is the time of closest approach to $\Gamma$, at which the trajectory ends.
We evaluate the minimum over the discrete states of the numerically propagated halo orbit, whose time step of at most \SI{e-3}{\TU} keeps the spacing below the reported arrival errors, with a pruned $k$-d tree search~\cite{Bentley.1975}.
\begin{table}[t!]
\caption{Mean costate and arrival errors of the final HEO-to-$\mathcal{L}_1$-halo guidance network (Sec.~\ref{sec: final guidance performance}) on \num{1000} initial states drawn from its held-out test split.
}
\label{tab:guidance_motivation}
\centering
\begin{tabular}{llS[table-format=1.4]}
\toprule
Metric & Evaluation & \multicolumn{1}{c}{Mean error} \\
\midrule
Costate error $\snorm{\hat{\bs{\lambda}}-\tilde{\bs{\lambda}}}$ & Direct prediction & \num{0.0023} \\
\midrule
\multirow{2}{*}{Arrival error $d_f$} & Open-loop guidance & \num{0.0226} \\
 & Guidance with $R=\num{30}$ replans & \num{0.0064} \\
\bottomrule
\end{tabular}
\end{table}

In principle, a single query suffices if the prediction is exact, because the returned costate then defines an extremal that reaches the target orbit.
However, trajectories in the CR3BP are highly sensitive to the initial costate, and the network predictions are not accurate enough for such open-loop use.
Table~\ref{tab:guidance_motivation} makes this concrete for the final HEO-to-$\mathcal{L}_1$-halo network, evaluated on initial states from its held-out test split, so that the states themselves lie within the training distribution.
The direct costate error of the first row is small, yet propagating that costate under the indirect dynamics turns it into the fairly large mean arrival error $d_f$ of the second row.
Replanning significantly reduces this error.
Querying the network again from the propagated state at each replan epoch cuts the mean arrival error by a factor of \num{3.5} relative to open-loop propagation.

\begin{algorithm}[b!]
\caption{Costate replanning guidance algorithm}
\label{alg:costate-replanning}
\algtext*{EndIf}%
\algtext*{EndFor}
\begin{algorithmic}[1]
\Require Initial state $\bs{x}_0$ at nominal time $t_0$; nominal trajectory; trained network; target orbit $\Gamma$; maximum replans $R$; arrival tolerance $\epsilon_a$; maximum refinement iterations $n_r$
\State $\bs{x}\gets\bs{x}_0$, $t\gets t_0$, $\textit{completed}\gets\emptyset$, and $\textit{best}\gets\emptyset$
\State $\Delta t_r\gets(t_{f,\nomsub}-t_0)/(R+1)$
\For{$i=0,\ldots,R$}
    \State $\textit{coast}\gets\Call{PropagateCoast}{\bs{x}}$
    \If{$\Call{ArrivalError}{\textit{coast},\Gamma}\leq\epsilon_a$}
        \State \Return $\textit{completed}$ followed by $\textit{coast}$
    \EndIf
    \State $\hat{\bs{\lambda}}\gets\Call{PredictCostate}{\bs{x},t}$
    \State $\textit{guided}\gets\Call{PropagateIndirect}{\bs{x},\hat{\bs{\lambda}}}$
    \If{$\Call{ArrivalError}{\textit{guided},\Gamma}>\epsilon_a$}
        \State $\textit{guided}\gets\Call{RefineCostate}{\textit{guided},n_r}$ \Comment{Appendix~\ref{sec: appendix costate refinement}}
    \EndIf
    \If{$\Call{ArrivalError}{\textit{guided},\Gamma}\leq\epsilon_a$}
        \State \Return $\textit{completed}$ followed by $\textit{guided}$
    \EndIf
    \State $\textit{plan}\gets\Call{BetterArrival}{\textit{coast},\textit{guided},\Gamma}$
    \State $\textit{candidate}\gets\textit{completed}$ followed by $\textit{plan}$
    \State $\textit{best}\gets\Call{BetterArrival}{\textit{best},\textit{candidate},\Gamma}$
    \If{$i=R$ \textbf{or} $t+\Delta t_r\geq t_{f,\nomsub}$}
        \State \Return $\textit{best}$
    \EndIf
    \State Append $\textit{plan}$ through $\Delta t_r$ to $\textit{completed}$
    \State $\bs{x}\gets\Call{StateAt}{\textit{plan},\Delta t_r}$ and $t\gets t+\Delta t_r$
\EndFor
\end{algorithmic}
\end{algorithm}

Algorithm~\ref{alg:costate-replanning} states the scheme in full, and we outline it here.
The algorithm divides the nominal time remaining at the initial state into $R+1$ equal replan intervals $\Delta t_r=(t_{f,\nomsub}-t_0)/(R+1)$, where $R$ is the maximum number of replans.
At each replan, \textsc{PropagateCoast} first integrates a ballistic coast over the mission-search window, which covers the remaining nominal flight time plus an additional quarter of the nominal time of flight.
The function \textsc{ArrivalError} then scores the coast trajectory by its closest arrival error $d_f$ from Eq.~\eqref{eq: arrival error}.
If that coast already satisfies $d_f\leq\epsilon_a$, where $\epsilon_a$ is the arrival tolerance, the algorithm returns it without querying the network.
This coast test is needed because the training dataset excludes terminal-coast samples, as described in Sec.~\ref{sec: costate training dataset}, so the network is never trained on a state from which the spacecraft reaches the orbit under zero thrust.
Otherwise, \textsc{PredictCostate} queries the guidance network at the current state and time, \textsc{PropagateIndirect} integrates the arc under the predicted costate over the same window, and \textsc{ArrivalError} determines its $d_f$.
If the guided arc misses, the optional \textsc{RefineCostate} corrects its costate with at most $n_r$ damped differential-correction iterations, as described in detail in Appendix~\ref{sec: appendix costate refinement}.
We evaluate the network both alone ($n_r=0$) and with this refinement.
If the guided arc arrives within $\epsilon_a$, the algorithm stops.
Otherwise, \textsc{BetterArrival} selects as the plan whichever of the coast and the guided arc arrives closer.
The trajectory completed so far, followed by the plan up to its closest approach, forms a candidate trajectory.
\textsc{BetterArrival} also keeps the best candidate over all replans, because a later plan can miss by more than an earlier one.
The algorithm then follows the plan for one replan interval $\Delta t_r$, and \textsc{StateAt} returns the state from which it replans.
After $R$ replans, the algorithm returns the best candidate.
The number of replans thus sets both the length of a guided segment and the number of network queries, and it trades runtime for accuracy (Sec.~\ref{sec: design studies}).
In our tests, an adaptive rule that replans where the guided arc passes closest to the nominal trajectory did not lower the arrival error compared with fixed segment lengths, and it adds the design choice of when to replan.
We use the arrival tolerance $\epsilon_a=\num{0.01}$ for the HEO-to-$\mathcal{L}_1$-halo transfer and $\epsilon_a=\num{0.02}$ for the HEO-to-$\mathcal{L}_2$-halo transfer.
Although fairly coarse, these tolerances were chosen such that they allow subsequent station keeping from the final state on the halo orbit using the LQR tracking law of Appendix~\ref{sec: appendix tracking baselines}.
This is verified in Appendix~\ref{sec: appendix station keeping}.

We assume that early costate predictions keep the subsequent states close to the dataset samples, so that later network queries remain within the training distribution.
This is reasonable because the training dataset contains multiple samples per trajectory and is constructed so that the initial sample is in distribution.
However, Sec.~\ref{sec: recovery behavior} shows that for the HEO-to-$\mathcal{L}_2$-halo transfer, small early prediction errors can make the trajectory diverge and push later queries out of distribution.

Because the costate jumps at each replan, the returned trajectory is not an extremal of the guidance problem itself.
Each guided segment, however, approximates an extremal of the free-final-time problem between the two boundary states it connects.
It satisfies the state and costate equations and the control law of Eq.~\eqref{eq: Control Law}, and the penalty in Eq.~\eqref{eq: training loss} drives its constant Hamiltonian toward $H=0$.
The refinement of Appendix~\ref{sec: appendix costate refinement} changes the Hamiltonian, so a refined segment can deviate further from this condition.

\section{Results}
\label{sec: results}
This section evaluates the guidance performance of the trained networks for both transfers.
We evaluate every network on the guidance test cases of Sec.~\ref{sec: guidance test scenarios} rather than on the test split held out from the training data, because these cases are generated separately from the training data and are more challenging and realistic.
We first present design studies for the HEO-to-$\mathcal{L}_1$-halo transfer, in which we compare training datasets, numbers of anchors, network sizes, and settings of the guidance algorithm, remove single components of the method, and select the final network.
We then compare our method with three baselines: constant-gain PD tracking and finite-horizon LQR tracking of the nominal trajectory, and re-solving the full nonconvex optimal control problem with SNOPT.
Finally, we examine the recovery behavior of the method on individual cases and discuss its limitations.

\subsection{Design Studies for the HEO-to-\texorpdfstring{$\mathcal{L}_1$}{L1}-Halo Transfer}
\label{sec: design studies}
\label{sec: dataset comparison}
\label{sec: model size comparison}
\begin{figure}[t!]
\centering
\includegraphics[width=\linewidth]{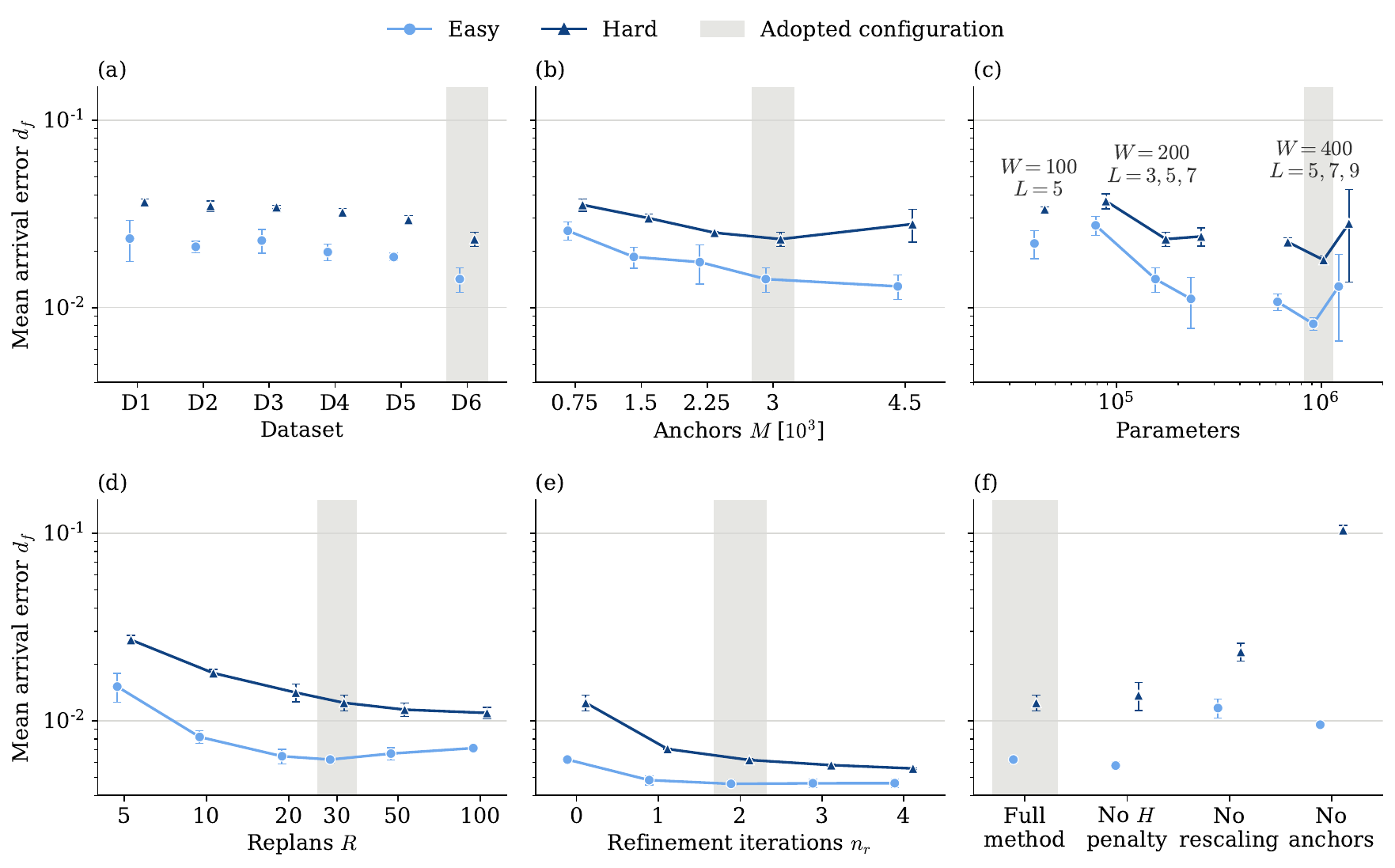}
\caption{Mean arrival error of the HEO-to-$\mathcal{L}_1$-halo design studies. Error bars show one standard deviation across three training seeds, and shading marks the adopted configuration.}
\label{fig:design-studies}
\end{figure}

Figure~\ref{fig:design-studies} summarizes the design studies for the HEO-to-$\mathcal{L}_1$-halo transfer on the Easy and Hard test cases, and Appendix~\ref{sec: appendix design studies} lists all values, including the Medium test case.
The metric is the mean arrival error $d_f$ of Eq.~\eqref{eq: arrival error} when the network is used within the guidance algorithm of Sec.~\ref{sec: guidance algorithm}, evaluated on \num{1000} samples from each of the three guidance test cases.
For each of the three training seeds, we average $d_f$ over these samples, and we report the mean and standard deviation of the three resulting values.
For the guidance settings, we also report the onboard cost as the mean number of floating-point operations (FLOPs) that the guidance algorithm executes per Hard trajectory, in millions (MFLOP).
Appendix~\ref{sec: appendix onboard computation} describes how we count them.
In Figs.~\ref{fig:design-studies}a--\ref{fig:design-studies}c, we select the training dataset, the number of anchors, and the network size with $R=\num{10}$ and $n_r=\num{0}$.
The first two studies use a network of $L=\num{5}$ layers of $W=\num{200}$ units.
In Figs.~\ref{fig:design-studies}d--\ref{fig:design-studies}f, we use the configuration adopted from these studies (D6, $W=\num{400}$, $L=\num{7}$) to study the guidance settings and the removal of single components of the method.

Figure~\ref{fig:design-studies}a compares the six datasets of Table~\ref{tab:dataset_design}, selected from a larger preliminary study.
Choosing $s$ per segment (D2) improves slightly on a uniform $s$ (D1), whereas doubling $s$ (D3) does not, likely because larger perturbations make the mapping harder to learn without covering the test cases better.
Mixing both perturbation sizes (D5) outperforms a single size with the same number of trajectories (D4) on every test case, and doubling $N$ helps for both recipes.
Dataset D4 outperforms D2, and D6 lowers the error of D5 by \SIrange{21}{33}{\percent}.
Larger $N$ may improve performance further, but each doubling also doubles the dataset size and the training time per epoch, so we adopt D6.

Figure~\ref{fig:design-studies}b shows that the error decreases as the number of anchors grows up to $M=\num{3000}$.
At $M=\num{4500}$, the Medium and Hard errors do not decrease further.
Their means rise, but the spread across training seeds more than doubles, so the results indicate a plateau rather than a clear degradation.
The additional anchors likely add samples mostly where the dataset is already dense, and anchors that converge to a different local optimum than their neighbors may make the mapping harder to learn, so we adopt $M=\num{3000}$.

Figure~\ref{fig:design-studies}c compares network widths $W$ and depths $L$ on dataset D6.
The largest gain comes from growing the smallest network to $W=\num{200}$, $L=\num{5}$, and in that range depth helps more per added parameter.
Going from $L=\num{3}$ to $L=\num{5}$ lowers the Hard error by \num{0.0139} with \num{80400} added parameters, whereas going from $W=\num{100}$ to $W=\num{200}$ lowers it by \num{0.0103} with \num{122000}.
Beyond that size, adding only depth ($W=\num{200}$, $L=\num{7}$) or only width ($W=\num{400}$, $L=\num{5}$) improves the Easy case but does not lower the Medium and Hard errors beyond the spread across training seeds.
Adding both gives the lowest error in every test case with the smallest spread across training seeds.
The network with $W=\num{400}$, $L=\num{7}$ lowers the Medium and Hard errors of $W=\num{200}$, $L=\num{5}$ by \SI{18}{\percent} and \SI{22}{\percent}.
The validation loss of these networks, shown in Appendix~\ref{sec: appendix design studies}, follows the same trend.
Up to $W=\num{400}$, $L=\num{7}$, larger networks generally reach a lower validation loss, and ranked by validation loss the architectures fall in nearly the order of their Hard errors.
The deepest network, $W=\num{400}$ with $L=\num{9}$, does not improve further.
Its validation loss is higher than that of $L=\num{7}$ for all three training seeds.
For one of these seeds the validation loss is lowest after only seven epochs, so the kept network comes from early in training, and its Hard error is \num{2.5} times the mean of $L=\num{7}$.
The networks of the other two seeds guide no better than those with $L=\num{7}$.
Because the network is intended for onboard use (Appendix~\ref{sec: appendix onboard computation}), we do not test larger networks and adopt $W=\num{400}$ and $L=\num{7}$.

Figure~\ref{fig:design-studies}d shows the effect of the number of replans $R$ without costate refinement ($n_r=\num{0}$).
Replanning more often lowers the error substantially at first.
From $R=\num{5}$ to $R=\num{30}$, the Medium and Hard errors fall by \SI{65}{\percent} and \SI{54}{\percent}.
Each replan also adds onboard computation, and over the same range the mean cost of a Hard trajectory grows from \SIrange{14.2}{47.2}{\mega\flop} (Appendix~\ref{sec: appendix design studies}).
Beyond $R=\num{30}$, the Easy error rises slightly, and up to $R=\num{100}$ the Medium and Hard errors fall by only \SI{2}{\percent} and \SI{12}{\percent}, whereas the cost grows to \SI{112.9}{\mega\flop}, so we adopt $R=\num{30}$.

Figure~\ref{fig:design-studies}e shows the effect of the maximum number of refinement iterations $n_r$ at $R=\num{30}$.
As with $R$, the error falls with diminishing returns as iterations are added, and every iteration adds onboard computation.
The first iteration gives most of the gain.
It lowers the Hard error by \SI{43}{\percent} and raises the mean cost of a Hard trajectory from \SIrange{47.2}{121.9}{\mega\flop}.
The second lowers the Hard error by another \SI{13}{\percent} for \SI{23.4}{\mega\flop} more.
Further iterations barely change the Easy and Medium errors and lower the Hard error by only \SI{6}{\percent} and \SI{3}{\percent}.
We therefore adopt $n_r=\num{2}$, and Sec.~\ref{sec: final guidance performance} reports the network both with this refinement and without it ($n_r=\num{0}$).

Finally, in Fig.~\ref{fig:design-studies}f, we remove one component of the method at a time from the adopted configuration: the anchor trajectories, the costate rescaling, or the Hamiltonian penalty.
We fly these networks without refinement ($n_r=\num{0}$), because refinement would partly correct the errors of a weaker network and hide the effect of the removed component.
The anchor trajectories matter most.
Without them, the network is trained on the same number of trajectories, generated with the same perturbations but around the nominal trajectory only.
Its Easy error rises by \SI{53}{\percent}, but its Medium and Hard errors grow by factors of \num{9.9} and \num{8.3}, because the outages of these test cases carry the spacecraft far from the nominal trajectory, where such a dataset has few samples.
Removing the costate rescaling has the second-largest effect and roughly doubles the error in every test case.
The Hamiltonian penalty has the smallest effect.
Removing it raises the Medium and Hard errors by \SI{8}{\percent} and \SI{10}{\percent} and lowers the Easy error by \SI{6}{\percent}, differences comparable to the spread across training seeds.

For the HEO-to-$\mathcal{L}_2$-halo transfer, Appendix~\ref{sec: appendix design studies} summarizes a smaller design study, which is less conclusive.
For datasets D1 to D5 at the same network size and number of replans, the errors of this transfer are five to seven times larger, and the spread across networks trained with different seeds and anchor partitions exceeds the differences between datasets, numbers of anchors, and network sizes.
The arrival error also follows the offline fit less closely.
Unlike on the HEO-to-$\mathcal{L}_1$-halo transfer, the larger network fits the validation data better without lowering the Hard error beyond the spread (Appendix~\ref{sec: appendix design studies}).
A likely reason is that the target halo orbit is strongly unstable, so small prediction errors early in the transfer grow until the network is queried outside its training distribution, and the arrival error becomes sensitive to small differences between networks.
Larger datasets or networks are therefore unlikely to help much on this transfer until later network queries are kept closer to the training data, a limitation that Sec.~\ref{sec: recovery behavior} returns to.

\subsection{Comparison with Guidance Baselines}
\label{sec: final guidance performance}
This subsection compares the final configuration of each transfer with three guidance baselines.
For the HEO-to-$\mathcal{L}_1$-halo transfer, we use dataset D6 with $M=\num{3000}$ anchors and a network with $W=\num{400}$ and $L=\num{7}$. 
For the HEO-to-$\mathcal{L}_2$-halo transfer, we use dataset D2 with $M=\num{1500}$ anchors and a network with $W=\num{200}$ and $L=\num{5}$.
Instead of averaging over three training seeds as in the preceding subsection, we report results for the seed with the lowest validation regression loss.
We run each network in the guidance algorithm with $R=\num{30}$ replans, once without refinement ($n_r=\num{0}$) and once with at most $n_r=\num{2}$ refinement iterations (Appendix~\ref{sec: appendix costate refinement}).
The first two baselines track the nominal trajectory, one with a PD law with constant gains and one with a finite-horizon LQR.
Appendix~\ref{sec: appendix tracking baselines} describes both and the selection of their gains.
The third baseline re-solves the full nonconvex fuel-optimal transfer problem from the off-nominal state to the target orbit with SNOPT, set up as for the anchor trajectories in Sec.~\ref{sec: anchor trajectories}.
Its feasibility tolerance is chosen so that every converged solution arrives within $\epsilon_a$.
It runs the solver twice with a time limit of \SI{120}{\second}, once from the costate of the nominal trajectory at the same epoch and once from the costate of the nearest anchor, and keeps the better result.
For each method, Tables~\ref{tab:l1_comparison} and~\ref{tab:l2_comparison} report the share of samples on which the spacecraft arrives within $\epsilon_a$, the mean and 95th percentile of the arrival error $d_f$, the fuel and time of flight (TOF) added relative to the nominal trajectory, and the onboard computation, counted as described in Appendix~\ref{sec: appendix onboard computation}.
\begin{table}[b!]
\caption{Network and baselines on the HEO-to-$\mathcal{L}_1$-halo transfer ($\epsilon_a=\num{0.01}$, $n_r=\num{2}$). Added fuel and TOF are means over the arrivals under each method and, in parentheses, over the samples on which the spacecraft arrives under every method (\num{882}, \num{452}, and \num{279} for Easy, Medium, and Hard). Boldface marks the best parenthesized value.}
\label{tab:l1_comparison}
\centering
\setlength{\tabcolsep}{5pt}
\begin{tabular}{lS[table-format=3.1]S[table-format=1.2e-1]S[table-format=1.2e-1]S[table-format=-2.1]@{\hspace{2pt}}rS[table-format=-2.1]@{\hspace{2pt}}rS[table-format=4.1]S[table-format=3.3]}
\toprule
& & \multicolumn{2}{c}{Arrival error $d_f$} & & & & & \multicolumn{2}{c}{Onboard computation\textsuperscript{a}} \\
\cmidrule(lr){3-4}\cmidrule(lr){9-10}
Method & \multicolumn{1}{c}{\makecell{Arrived\\{[\si{\percent}]}}} & \multicolumn{1}{c}{Mean} & \multicolumn{1}{c}{95th pct.} & \multicolumn{2}{c}{\makecell{Added\\fuel [\si{\percent}]}} & \multicolumn{2}{c}{\makecell{Added\\TOF [\si{\percent}]}} & \multicolumn{1}{c}{\makecell{Operations\\{[\si{\mega\flop}]}}} & \multicolumn{1}{c}{\makecell{Time\\{[\si{\second}]}}} \\
\midrule
\multicolumn{10}{l}{\textit{Easy}} \\
Network & \num{98.8} & \num{5.91e-3} & \num{9.63e-3} & \num{0.2} & (\num{0.1}) & \num{-0.1} & (\bestnum{-0.1}) & \num{26.6} & \num{1.332} \\
Network + refine & \num{99.8} & \num{4.53e-3} & \num{9.18e-3} & \num{0.2} & (\num{0.1}) & \num{-0.2} & (\bestnum{-0.1}) & \num{99.4} & \num{4.968} \\
PD tracking & \bestnum{100.0} & \bestnum{1.86e-5} & \bestnum{5.98e-5} & \num{1.0} & (\num{1.0}) & \num{0.0} & (\num{0.0}) & \bestnum{7.5} & \bestnum{0.377} \\
LQR tracking & \bestnum{100.0} & \num{2.20e-4} & \num{5.27e-4} & \num{0.3} & (\num{0.3}) & \num{1.7} & (\num{1.7}) & \num{7.6} & \num{0.382} \\
Re-solve (SNOPT) & \num{89.1} & \num{1.83e-2} & \num{1.50e-1} & \num{-0.9} & (\bestnum{-1.0}) & \num{6.5} & (\num{6.6}) & \num{6760.4} & \num{338.021} \\
\midrule
\multicolumn{10}{l}{\textit{Medium}} \\
Network & \num{85.7} & \num{8.24e-3} & \num{2.54e-2} & \num{7.2} & (\num{5.1}) & \num{0.0} & (\num{-0.2}) & \num{38.5} & \num{1.924} \\
Network + refine & \bestnum{99.2} & \bestnum{4.79e-3} & \bestnum{9.34e-3} & \num{8.4} & (\num{4.9}) & \num{-0.5} & (\bestnum{-0.3}) & \num{87.6} & \num{4.380} \\
PD tracking & \num{86.2} & \num{1.61e-1} & \num{1.21e0} & \num{19.3} & (\num{12.4}) & \num{0.1} & (\num{0.0}) & \bestnum{6.6} & \bestnum{0.332} \\
LQR tracking & \num{73.1} & \num{9.58e-3} & \num{2.33e-2} & \num{10.9} & (\num{6.5}) & \num{9.3} & (\num{8.9}) & \num{6.7} & \num{0.337} \\
Re-solve (SNOPT) & \num{75.7} & \num{9.67e-2} & \num{6.04e-1} & \num{7.7} & (\bestnum{4.5}) & \num{5.8} & (\num{6.1}) & \num{8926.1} & \num{446.304} \\
\midrule
\multicolumn{10}{l}{\textit{Hard}} \\
Network & \num{73.2} & \num{1.23e-2} & \num{4.06e-2} & \num{11.0} & (\num{5.5}) & \num{0.3} & (\num{-0.3}) & \num{47.2} & \num{2.360} \\
Network + refine & \bestnum{94.4} & \bestnum{6.33e-3} & \bestnum{1.20e-2} & \num{13.6} & (\num{5.1}) & \num{-0.9} & (\bestnum{-0.4}) & \num{145.3} & \num{7.267} \\
PD tracking & \num{72.4} & \num{2.55e-1} & \num{1.28e0} & \num{23.9} & (\num{13.3}) & \num{1.6} & (\num{0.0}) & \bestnum{6.3} & \bestnum{0.314} \\
LQR tracking & \num{51.2} & \num{7.61e-2} & \num{6.00e-2} & \num{11.9} & (\num{6.7}) & \num{9.6} & (\num{9.1}) & \num{6.4} & \num{0.318} \\
Re-solve (SNOPT) & \num{76.9} & \num{7.41e-2} & \num{4.15e-1} & \num{14.1} & (\bestnum{4.9}) & \num{4.8} & (\num{5.2}) & \num{8766.1} & \num{438.307} \\
\bottomrule
\end{tabular}
\par\smallskip
{\raggedright\textsuperscript{a}Mean per trajectory; time at \SI[per-mode=symbol]{20}{\mega\flop\per\second}, about what one core of the GR740 processor sustains (Appendix~\ref{sec: appendix onboard computation}).\par}
\end{table}

\begin{table}[t!]
\caption{
Network and baselines on the HEO-to-$\mathcal{L}_2$-halo transfer ($\epsilon_a=\num{0.02}$, $n_r=\num{2}$). 
Added fuel and TOF are means over the arrivals under each method and, in parentheses, over the samples on which the spacecraft arrives under every method (\num{317}, \num{188}, and \num{112} for Easy, Medium, and Hard). Boldface marks the best parenthesized value.
}
\label{tab:l2_comparison}
\centering
\setlength{\tabcolsep}{5pt}
\begin{tabular}{lS[table-format=3.1]S[table-format=1.2e-1]S[table-format=1.2e-1]S[table-format=-2.1]@{\hspace{2pt}}rS[table-format=-2.1]@{\hspace{2pt}}rS[table-format=5.1]S[table-format=3.3]}
\toprule
& & \multicolumn{2}{c}{Arrival error $d_f$} & & & & & \multicolumn{2}{c}{Onboard computation} \\
\cmidrule(lr){3-4}\cmidrule(lr){9-10}
Method & \multicolumn{1}{c}{\makecell{Arrived\\{[\si{\percent}]}}} & \multicolumn{1}{c}{Mean} & \multicolumn{1}{c}{95th pct.} & \multicolumn{2}{c}{\makecell{Added\\fuel [\si{\percent}]}} & \multicolumn{2}{c}{\makecell{Added\\TOF [\si{\percent}]}} & \multicolumn{1}{c}{\makecell{Operations\\{[\si{\mega\flop}]}}} & \multicolumn{1}{c}{\makecell{Time\\{[\si{\second}]}}} \\
\midrule
\multicolumn{10}{l}{\textit{Easy}} \\
Network & \num{60.0} & \num{5.74e-2} & \num{3.62e-1} & \num{5.1} & (\num{3.9}) & \num{1.4} & (\num{1.3}) & \num{114.6} & \num{5.729} \\
Network + refine & \num{92.5} & \num{2.17e-2} & \num{5.69e-2} & \num{16.1} & (\num{7.7}) & \num{-2.6} & (\num{1.9}) & \num{579.9} & \num{28.994} \\
PD tracking & \bestnum{100.0} & \bestnum{8.83e-5} & \bestnum{2.79e-4} & \num{0.4} & (\num{0.3}) & \num{0.0} & (\bestnum{0.0}) & \bestnum{7.4} & \bestnum{0.369} \\
LQR tracking & \bestnum{100.0} & \num{6.25e-4} & \num{1.82e-3} & \num{0.2} & (\bestnum{0.2}) & \num{2.4} & (\num{2.6}) & \bestnum{7.4} & \num{0.372} \\
Re-solve (SNOPT) & \num{46.6} & \num{1.82e-1} & \num{6.48e-1} & \num{2.2} & (\num{1.0}) & \num{2.5} & (\num{2.3}) & \num{13751.1} & \num{687.557} \\
\midrule
\multicolumn{10}{l}{\textit{Medium}} \\
Network & \num{40.2} & \num{1.26e-1} & \num{4.66e-1} & \num{10.7} & (\num{6.7}) & \num{1.9} & (\bestnum{1.3}) & \num{110.3} & \num{5.516} \\
Network + refine & \bestnum{72.8} & \num{6.53e-2} & \num{3.50e-1} & \num{17.0} & (\num{6.2}) & \num{2.0} & (\num{4.2}) & \num{551.6} & \num{27.579} \\
PD tracking & \num{68.0} & \num{2.71e-1} & \num{1.18e0} & \num{11.0} & (\num{5.9}) & \num{1.1} & (\num{1.6}) & \bestnum{6.9} & \bestnum{0.346} \\
LQR tracking & \num{60.6} & \bestnum{3.71e-2} & \bestnum{1.55e-1} & \num{7.9} & (\num{3.9}) & \num{3.5} & (\num{3.3}) & \num{7.0} & \num{0.348} \\
Re-solve (SNOPT) & \num{53.7} & \num{1.75e-1} & \num{6.93e-1} & \num{10.3} & (\bestnum{3.6}) & \num{3.1} & (\num{2.5}) & \num{17825.7} & \num{891.283} \\
\midrule
\multicolumn{10}{l}{\textit{Hard}} \\
Network & \num{34.8} & \num{1.56e-1} & \num{5.11e-1} & \num{13.1} & (\num{7.3}) & \num{1.7} & (\bestnum{1.3}) & \num{109.5} & \num{5.477} \\
Network + refine & \bestnum{68.2} & \bestnum{8.02e-2} & \bestnum{4.41e-1} & \num{20.5} & (\num{6.9}) & \num{0.9} & (\num{4.9}) & \num{535.3} & \num{26.766} \\
PD tracking & \num{50.6} & \num{3.76e-1} & \num{1.22e0} & \num{12.6} & (\num{7.1}) & \num{1.7} & (\num{2.4}) & \bestnum{6.7} & \bestnum{0.334} \\
LQR tracking & \num{42.6} & \num{1.42e-1} & \num{9.22e-1} & \num{8.5} & (\num{4.4}) & \num{4.2} & (\num{3.1}) & \bestnum{6.7} & \num{0.337} \\
Re-solve (SNOPT) & \num{50.0} & \num{1.64e-1} & \num{6.40e-1} & \num{14.2} & (\bestnum{4.2}) & \num{4.6} & (\num{2.4}) & \num{19725.0} & \num{986.249} \\
\bottomrule
\end{tabular}
\end{table}

The arrival percentages in Tables~\ref{tab:l1_comparison} and~\ref{tab:l2_comparison} demonstrate that the spacecraft reaches the target orbit more often under our method than under any baseline for the Medium and Hard cases on both transfers. 
On the Easy cases, the spacecraft arrives at the target orbit within tolerance on every sample under PD and LQR tracking. 
Intuitively, this is because for the Easy case, the off-nominal trajectories still start sufficiently close to the nominal trajectory, and hence the relative model is appropriate on the timescales considered. 
Tracking works less well the farther an outage carries the spacecraft from that nominal trajectory, and under PD tracking the spacecraft misses the target orbit on \SI{27.6}{\percent} and \SI{49.4}{\percent} of the Hard samples of the HEO-to-$\mathcal{L}_1$-halo and HEO-to-$\mathcal{L}_2$-halo transfers.

On the HEO-to-$\mathcal{L}_1$-halo transfer, the network without refinement already matches PD tracking on the Medium and Hard cases, and refinement raises its Hard arrival rate to \SI{94.4}{\percent}, against \SI{76.9}{\percent} for the best baseline, which is the re-solve with SNOPT.
On the HEO-to-$\mathcal{L}_2$-halo transfer, the network without refinement falls short of every baseline on the Medium and Hard cases, but with refinement, the spacecraft again arrives most often under network guidance, on \SI{68.2}{\percent} of the Hard samples against \SI{50.6}{\percent} for the best baseline.
With the re-solve, the spacecraft arrives on only about half of the HEO-to-$\mathcal{L}_2$-halo samples, even in the Easy case, which shows how sensitive the problem is to the initial guess.
SNOPT often does not converge from either the nominal or the nearest-anchor costate.

Another advantage of our approach compared with the baseline methods is that when the spacecraft misses under network guidance, it usually misses more narrowly than under PD tracking and the re-solve, with which it often ends far from the target orbit.
The $d_f$ metrics in Tables~\ref{tab:l1_comparison} and~\ref{tab:l2_comparison} show this behavior.
The low mean and 95th percentile values for LQR and PD tracking on the Easy cases demonstrate that these methods bring the spacecraft within much tighter tolerances than required.
However, for the Medium and Hard cases, a sharp increase in those values for PD tracking follows from samples on which the spacecraft does not even reach the vicinity of the target orbit.
Under LQR tracking, the spacecraft misses more often than under PD tracking, but only a small share of these misses end far from the target orbit.
Network guidance also shows no such increase, and with refinement, it achieves lower values on both $d_f$ metrics than every baseline for the Medium and Hard HEO-to-$\mathcal{L}_1$-halo cases.
Notably, the network with refinement beats all baseline methods by at least an order of magnitude in terms of mean $d_f$ on the Hard HEO-to-$\mathcal{L}_1$-halo case.
On the HEO-to-$\mathcal{L}_2$-halo transfer, it has the lowest mean and 95th percentile of $d_f$ on the Hard case, whereas LQR tracking has the lowest on the Medium case.

In terms of fuel consumption and time of flight, our approach is more efficient than the tracking baselines on the HEO-to-$\mathcal{L}_1$-halo transfer.
On the samples on which the spacecraft arrives under every method (in parentheses), network guidance requires less additional fuel than PD and LQR tracking in all cases of this transfer, both with and without refinement, and its time of flight stays within \SI{0.5}{\percent} of the nominal one, whereas LQR tracking adds about \SI{9}{\percent} on the Medium and Hard cases.
PD tracking requires by far the most additional fuel.
This behavior is expected.
Tracking forces the spacecraft back onto the nominal trajectory instead of replanning from its current state, and PD tracking, unlike LQR tracking, does not weigh the control effort against the tracking error.
Because the network predicts the costates of fuel-optimal extremals, each guided arc follows the fuel-optimal control law, so a low fuel cost is expected.
The re-solve still uses the least fuel in all cases of this transfer, but on the Medium and Hard cases only \num{0.2} to \num{0.4} percentage points less than the network with refinement, and its trajectories take \SIrange{5}{7}{\percent} longer than the nominal trajectory.
On the HEO-to-$\mathcal{L}_2$-halo transfer, the picture is different.
The network needs more fuel than LQR tracking and the re-solve in every case, about as much as PD tracking on the Medium and Hard cases, and much more than every baseline on the Easy case.
This again points to the limitations of our approach on this more sensitive transfer.

\begin{figure}[t!]
\centering
\includegraphics[width=\linewidth]{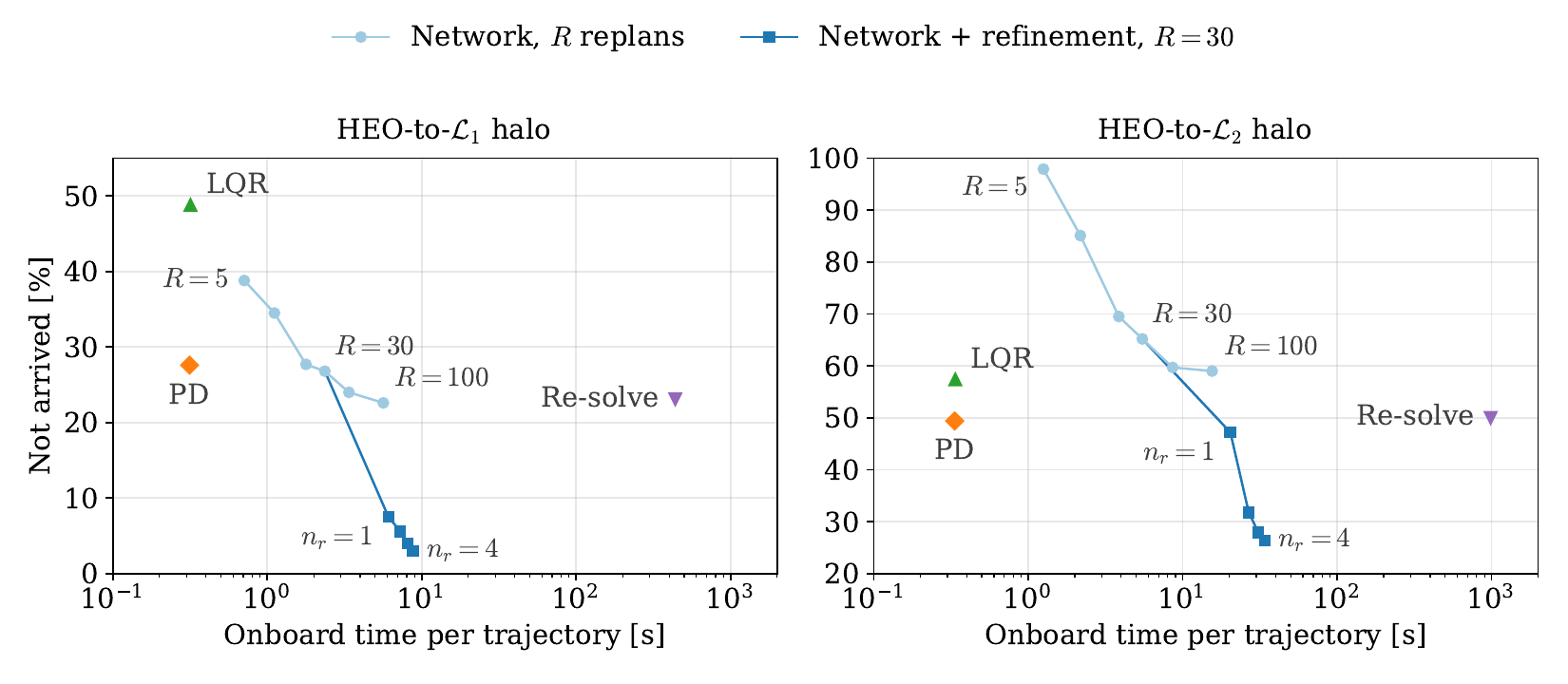}
\caption{Share of Hard test samples without arrival within $\epsilon_a$ versus onboard time per trajectory, for the network with $R$ replans or refinement and the baselines.}
\label{fig:arrival-vs-onboard-cost}
\end{figure}

Overall, our method provides a favorable trade-off between arrival rate and computational cost on the Hard cases of both transfers. 
For onboard applications, the number of FLOPs performed during flight and the associated computation time are important performance factors.
As expected, the re-solve is by far the most expensive method.
It needs more than an order of magnitude more FLOPs than any other method, about \SIrange{6}{16}{\minute} per trajectory at the rate of Appendix~\ref{sec: appendix onboard computation}.
The two tracking methods are the cheapest, followed by the network without refinement.
Runtime does not limit our method.
At the same rate, its whole guidance loop takes at most about half a minute per trajectory, even with refinement, whereas a replan with the ground in the loop can take days.
Figure~\ref{fig:arrival-vs-onboard-cost} shows the full trade-off on the Hard cases of both transfers, including the network without refinement at several numbers of replans $R$ and, at $R=\num{30}$, with up to four refinement iterations $n_r$.
On both transfers, PD tracking and the network with one to four refinement iterations lie on the Pareto front of this trade-off, whereas LQR tracking and the re-solve are dominated.
PD tracking has a higher arrival rate than LQR tracking at about the same cost, and every refinement point has a higher arrival rate than the re-solve at a small fraction of the re-solve's cost.
On the HEO-to-$\mathcal{L}_1$-halo transfer, the network without refinement also lies on the front for $R\geq\num{30}$, because its arrival rate exceeds that of PD tracking at less cost than any refinement point.
Which point to choose depends on the computation available on board, the arrival rate the mission requires, and the added fuel and time of flight it can accept.
Varying the number of refinement iterations moves the operating point of one trained network along the front without retraining.

\subsection{Recovery Behavior and Limitations}
\label{sec: recovery behavior}
\begin{figure}[t!]
\centering
\includegraphics[width=\linewidth]{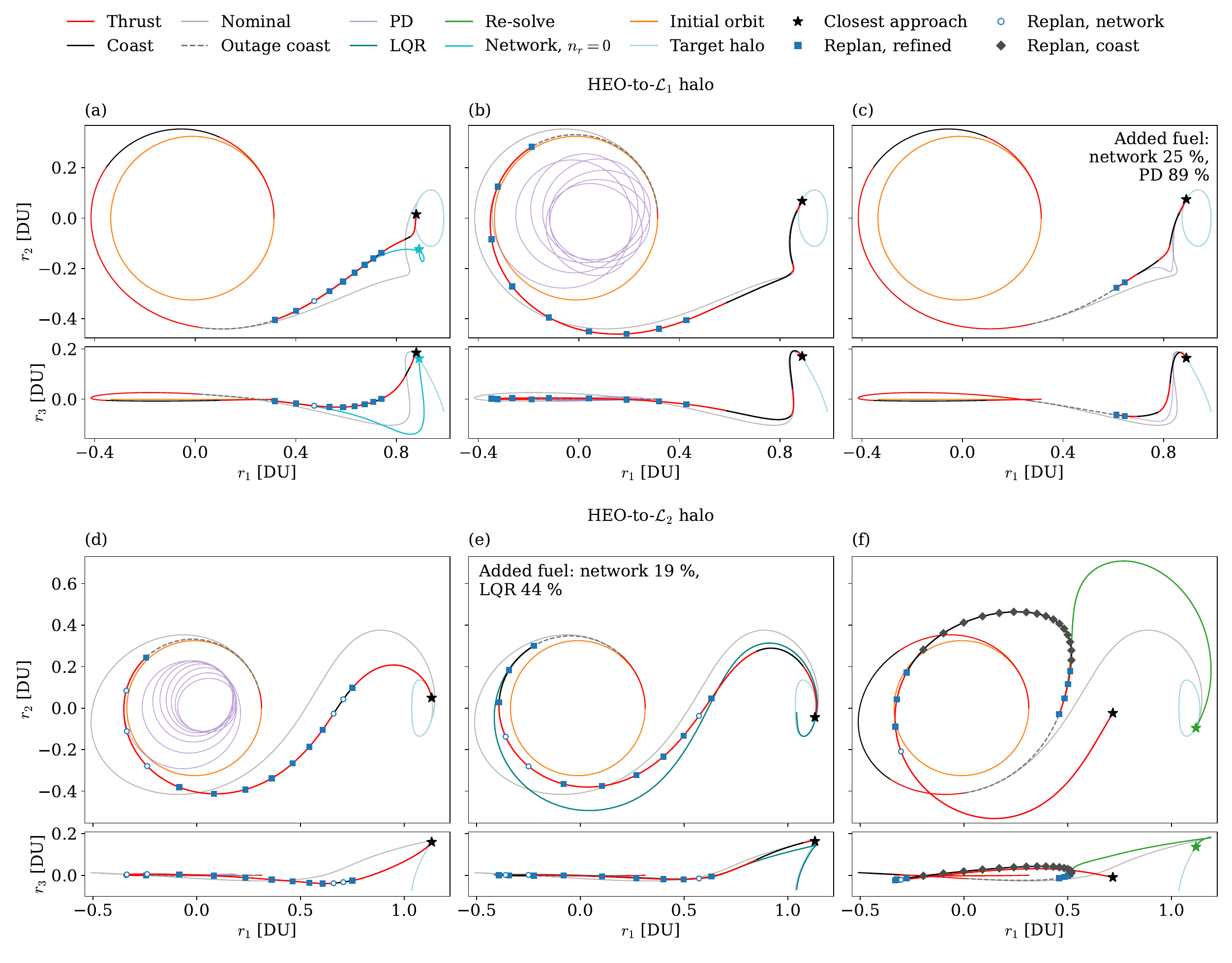}
\caption{Hard-case recoveries of the network with refinement and of the baselines on the HEO-to-$\mathcal{L}_1$-halo (top) and HEO-to-$\mathcal{L}_2$-halo (bottom) transfers.}
\label{fig:recovery-examples}
\end{figure}

This subsection uses example trajectories to show how the spacecraft recovers under our method, where the method improves on the baselines, and where it fails, and then discusses the limitations these examples point to.
Figure~\ref{fig:recovery-examples} shows six Hard recoveries, three per transfer, of the network with $R=\num{30}$ and $n_r=\num{2}$, the configuration of Tables~\ref{tab:l1_comparison} and~\ref{tab:l2_comparison}.
Each panel compares it with one alternative: the network without refinement (a), PD tracking (b--d), LQR tracking (e), and the re-solve (f).
The markers on each flown trajectory are its replans and distinguish plans taken directly from the network, refined plans, and coasts.

Figure~\ref{fig:recovery-examples}a shows how refinement turns a miss into an arrival.
After this outage in the second thrust segment of the HEO-to-$\mathcal{L}_1$-halo transfer, the spacecraft misses the target orbit under the network alone ($d_f=\num{0.116}$).
With refinement, eight of the first nine flown arcs are refined, and the plan of the ninth replan arrives at $d_f=\num{0.0083}$.

Figures~\ref{fig:recovery-examples}b and~\ref{fig:recovery-examples}d show outages early in the first thrust segment, after which the spacecraft misses the target orbit under PD tracking but arrives under our method.
In those cases, the PD feedback law is not able to track the nominal trajectory, and the spacecraft instead spirals toward the Earth, leading to very large arrival errors ($d_f=\num{1.29}$ and \num{1.35}).
Under LQR tracking, which is not shown in the figure, the spacecraft misses in those cases in the same way ($d_f=\num{1.70}$ and \num{1.35}).
Network guidance instead replans the remaining transfer, along a trajectory that stays close to the nominal one in Fig.~\ref{fig:recovery-examples}b and takes a markedly different route in Fig.~\ref{fig:recovery-examples}d.
The spacecraft arrives along the plans of the 9th and 14th replans ($d_f=\num{0.0063}$ and \num{0.0022}).

Figures~\ref{fig:recovery-examples}c and~\ref{fig:recovery-examples}e show samples on which the spacecraft also arrives under a tracking law, but at a much higher cost.
The trackers command thrust throughout, PD at full throttle for \SI{82}{\percent} of the flight in Fig.~\ref{fig:recovery-examples}c and LQR for \SI{69}{\percent} in Fig.~\ref{fig:recovery-examples}e, whereas the network-guided trajectory keeps a thrust--coast structure close to the nominal one.
The added fuel, defined as in Tables~\ref{tab:l1_comparison} and~\ref{tab:l2_comparison}, is \SI{25}{\percent} for the network against \SI{89}{\percent} for PD tracking, and \SI{19}{\percent} against \SI{44}{\percent} for LQR tracking.

Figure~\ref{fig:recovery-examples}f shows a failure case for our method, which occurs after an outage late in the second thrust segment of the HEO-to-$\mathcal{L}_2$-halo transfer.
The state after the outage is within the training distribution, but near its edge.
Only \SI{5}{\percent} of the held-out validation samples are farther from their nearest training sample.
Because this state lies within the most sensitive part of the transfer, the network's first plan with refinement misses the target orbit by \num{0.59}.
After five replans, the trajectory is far from all training samples, so the network's predictions are of little use, and the spacecraft makes another revolution around the Earth.
Both tracking methods also fail in this case, whereas with the re-solve, the spacecraft reaches the target orbit along a trajectory that departs far from the nominal one, a recovery unlike the trajectories the network was trained on.

We hypothesize that the failure mechanism for the case shown in Fig. ~\ref{fig:recovery-examples}f is the same that explains many of the other misses for both halo transfers. 
Adding anchors along the states that the closed loop visits could help keep later queries inside the distribution, similar to the dataset aggregation approach of Ross et al.~\cite{Ross.2011}.
A miss also need not end the mission because the whole loop runs in simulation before the spacecraft fully executes it. 
A miss is known before any thrust is executed, and the guidance system can fall back to another more expensive method or to a replan with the ground in the loop.
Still, onboard use would benefit from arrival rates closer to \SI{100}{\percent}, so raising the Hard arrival rates remains an important goal of future work.

\section{Conclusion}
\label{sec: conclusion}
This paper has presented a deep learning method, built on three important contributions, that enables the generation of new low-thrust trajectories for autonomous onboard spacecraft replanning in a multibody system for the case where the true trajectory has deviated far from its nominal path.
The first contribution is that the mathematical formulation of BGOE has been extended to the CR3BP.
The second contribution demonstrated that adding anchor trajectories to the data generation and rescaling the training costates are essential for guidance performance.
A loss term was also added to the training phase of the network, and although enabling improvements, is of lesser impact than the anchor trajectory data augmentation approach and rescaling of costate data. 
The third contribution combined repeated network replans with costate refinement in a guidance algorithm that achieves the highest arrival rate of all methods on the Medium and Hard test cases of both Earth--Moon transfers.
Additional contributions of this paper include the detailed network hyperparameter and architectural studies and guidance on filtering of data for effective training of the network. 

On the HEO-to-$\mathcal{L}_1$-halo transfer, the spacecraft reaches the target orbit on \SI{94}{\percent} of the Hard samples with our method, compared with \SI{72}{\percent} and \SI{51}{\percent} with PD and LQR tracking of the nominal trajectory.
Our method also yields more arrivals than re-solving the full nonconvex optimal control problem with SNOPT, and does so with one-sixtieth of the computation cost and a similar fuel usage.
Although the method yields impressive improvements in both transfers considered, the arrival errors remain an order of magnitude larger on the Medium and Hard cases of the HEO-to-$\mathcal{L}_2$-halo transfer than on the HEO-to-$\mathcal{L}_1$-halo transfer.
Improving performance on the more difficult transfer through the circumlunar region to a strongly unstable $\mathcal{L}_2$ halo orbit is the primary direction for future work.

Several directions are ongoing or should be pursued for additional advances of the method proposed in this manuscript. 
The evaluation in this paper is deterministic.
In each test sample, the state is perturbed once, and the network's plans are then flown with perfect state knowledge and without execution errors.
A flight implementation would face these errors continuously, so future work includes closing the GNC loop and accounting for these uncertainties. 
In particular, we plan to develop uncertainty quantification results that provide informative bounds on the sensitivity of the trained network to these uncertainties and to the variability of the training dataset.
Extending BGOE to check second-order conditions could further improve the training data by excluding extremals that are not locally optimal.
Finally, the method should be extended and tested in higher-fidelity dynamical models to understand generalizability.

\appendixsection{Costate Refinement}
\label{sec: appendix costate refinement}
The costate refinement \textsc{RefineCostate} of Algorithm~\ref{alg:costate-replanning} corrects the predicted costate $\hat{\bs{\lambda}}$ of a guided arc that misses the target orbit, $d_f>\epsilon_a$.
Let the arc from the replan state $\bs{x}$ at time $t$ make its closest approach to $\Gamma$ at time $t_f$, at the target-orbit state with final coast time $\tau_f$, and let the arrival miss $\bs{d}\in\mathbb{R}^6$ be the vector inside the norm of Eq.~\eqref{eq: arrival error} at this minimum, so that $d_f=\snorm{\bs{d}}$.
The refinement step computes a costate change $\delta\bs{\lambda}$, a shift $\delta t$ of the arrival time, and a shift $\delta\tau_f$ of the arrival point along $\Gamma$ as the damped least-squares (Levenberg--Marquardt) solution of
\begin{equation}
\min_{\delta\bs{\lambda},\delta t,\delta\tau_f}
\norm{\bs{d}+\bs{\Phi}\delta\bs{\lambda}+\frac{\partial\bs{d}}{\partial t_f}\delta t+\frac{\partial\bs{d}}{\partial\tau_f}\delta\tau_f}^2
+\gamma\snorm{\bs{\Phi}}_2^2\snorm{\delta\bs{\lambda}}^2
\label{eq: costate refinement step}
\end{equation}
whose first term is the miss linearized in these nine unknowns.
Here $\bs{\Phi}=\partial\bs{d}/\partial\bs{\lambda}(t)\in\mathbb{R}^{6\times7}$ is the position--velocity block of the state--costate transition matrix at fixed $t_f$, which \texttt{pydylan} computes from the variational equations across the thrust switches.
The other two derivatives are the rates of the arc at $t_f$ and of the target orbit at $\tau_f$, which follow from the controlled dynamics of Eq.~\eqref{eq: controlled low-thrust CR3BP dynamics}, with zero thrust on the orbit, as
\begin{equation*}
\renewcommand{\arraystretch}{0.6}%
\frac{\partial\bs{d}}{\partial t_f}=\begin{pmatrix}\bs{v}\\ \bs{g}(\bs{r},\bs{v})+\frac{T}{m}\hat{\bs{u}}\end{pmatrix},\qquad
\frac{\partial\bs{d}}{\partial\tau_f}=\begin{pmatrix}\bs{v}_\Gamma\\ \bs{g}(\bs{r}_\Gamma,\bs{v}_\Gamma)\end{pmatrix}
\end{equation*}
where the positive sign in the second equation follows from the definition of $\tau_f$. 
The damping, scaled by the squared largest singular value $\snorm{\bs{\Phi}}_2^2$ so that $\gamma$ is dimensionless, keeps the costate change small, because $\bs{\Phi}$ is poorly conditioned and the undamped step usually leaves the range of the linearization.
The damping also ensures that the step turns the costate rather than rescaling it, which would leave the trajectory unchanged.
Equation~\eqref{eq: costate refinement step} is solved through a singular-value decomposition.
Each iteration computes the step for $\gamma=\num{e-4}$, \num{e-3}, and \num{e-2} in turn, from the largest costate change to the smallest, and keeps the first corrected arc that arrives, or otherwise the closest of the three.
If the kept arc arrives closer than the arc being corrected, it replaces that arc, and if the kept arc still misses, the next iteration linearizes about it.
Arcs that reach their closest approach less than \SI{0.1}{\TU} after the replan are not refined, because little thrust time remains to change their miss.

The step corrects the miss but not the optimality of the arc, so a refined arc satisfies the free-final-time condition $H=0$ less closely than the network's prediction.
Adding the linearized condition $\delta H=0$ to the step did not improve the arrival rate or the fuel, and for the HEO-to-$\mathcal{L}_2$-halo transfer it made both worse, so we omit this condition from the refinement.
In our tests, the arc from the minimum-norm solution of Eq.~\eqref{eq: costate refinement step} without the damping term arrived less often than the damped arc, even when the step was shortened to half or a quarter of its length.

\appendixsection{Tracking Baselines}
\label{sec: appendix tracking baselines}
The LQR and PD tracking baselines both steer the spacecraft with the nominal thrust plus a linear feedback on the deviation from the nominal state at the same clock time, from the off-nominal state $\bs{x}_0$ at its nominal epoch $t_0$ to the nominal arrival time $t_{f,\nomsub}$.
Both command the thrust acceleration
\begin{equation*}
\bs{a}=\sigma_\nomsub\frac{T_{\max}}{m}\hat{\bs{u}}_\nomsub+\Delta\bs{a}
\end{equation*}
where $\sigma_\nomsub$ and $\hat{\bs{u}}_\nomsub$ are the nominal throttle and thrust direction, and $\Delta\bs{a}$ is the feedback.
The command is held constant over control intervals of $\Delta t_c=\SI{0.01}{\TU}$, with the interval average of the nominal control and the feedback on the position and velocity deviation $\Delta\bs{x}_{rv}=(\Delta\bs{r}\T,\Delta\bs{v}\T)\T$ at the interval start.
A command beyond the thrust bound keeps its direction and saturates the throttle, $\sigma=\min(m\snorm{\bs{a}}/T_{\max},1)$.
As for the network, a coast test at $\bs{x}_0$ with tolerance $\epsilon_a$ precedes the flight, and after $t_{f,\nomsub}$ the spacecraft coasts through the rest of the mission-search window, and the trajectory is scored by its arrival error $d_f$.
The gains of both laws were selected once per transfer on \num{200} tuning cases drawn with the Medium-case outage durations of Sec.~\ref{sec: guidance test scenarios} and a new seed, sharing no initial state with the test cases.
The gains with the lowest mean arrival error over the tuning cases in which the spacecraft arrives within $\epsilon_a$ for at least one setting of the sweep were selected.

\subsection*{Constant-Gain PD Tracking}
PD tracking applies constant gains to the position and velocity deviations,
\begin{equation*}
\Delta\bs{a}=-\omega_n^2\Delta\bs{r}-2\zeta\omega_n\Delta\bs{v}
\end{equation*}
so that the deviation behaves as a damped oscillator with natural frequency $\omega_n$ and damping ratio $\zeta$ if the gravity gradient and the Coriolis acceleration acting on it are neglected.
The sweep covered $\omega_n\in\{1,2,3,5,8,12,18,25,35\}$~\si{\per\TU}, well below the stability limit of the sampled loop, and $\zeta\in\{0.7,1.0,1.4,2.0,2.8\}$.
The selected gains are $\omega_n=\SI{12}{\per\TU}$ and $\zeta=\num{1.0}$ for the HEO-to-$\mathcal{L}_1$-halo transfer and $\omega_n=\SI{8}{\per\TU}$ and $\zeta=\num{1.4}$ for the HEO-to-$\mathcal{L}_2$-halo transfer.

\subsection*{Finite-Horizon LQR Tracking}
LQR tracking instead takes its gains from the deviation dynamics linearized about the nominal trajectory,
\begin{equation*}
\renewcommand{\arraystretch}{0.6}%
\Delta\dot{\bs{x}}_{rv} = \bs{A}(t)\Delta\bs{x}_{rv} + \bs{B}\Delta\bs{a}, \qquad
\bs{A}(t) = \begin{pmatrix} \bs{0} & \bs{I}_3 \\ \bs{G}_r(\bs{r}_\nomsub(t)) & \bs{G}_v \end{pmatrix}, \qquad
\bs{B} = \begin{pmatrix} \bs{0} \\ \bs{I}_3 \end{pmatrix}
\end{equation*}
with the Jacobians $\bs{G}_r$ and $\bs{G}_v$ of Eq.~\eqref{eq: Costate equation of motion}, and minimizes the quadratic cost
\begin{equation*}
J_\mathrm{LQR} = \frac{1}{2}\Delta\bs{x}_{rv}(t_{f,\nomsub})\T \bs{P}_f\Delta\bs{x}_{rv}(t_{f,\nomsub})
+ \frac{1}{2}\int_{t_0}^{t_{f,\nomsub}} \left( \Delta\bs{x}_{rv}\T \bs{Q}_x\Delta\bs{x}_{rv} + \Delta\bs{a}\T \bs{Q}_a\Delta\bs{a} \right) \, dt
\end{equation*}
Its feedback is $\Delta\bs{a}=-\bs{Q}_a^{-1}\bs{B}\T\bs{P}(t)\Delta\bs{x}_{rv}$, with $\bs{P}(t)$ from the Riccati equation integrated backward from $\bs{P}_f$ once per transfer.
The weights follow Bryson's rule~\cite{Bryson.1975}.
The state weight $\bs{Q}_x=\num{e4}\bs{I}_6$ penalizes deviations at the arrival-error scale of \num{0.01}, and the control weight $\bs{Q}_a=\kappa(m_0/T_{\max})^2\bs{I}_3$ scales the control by the thrust acceleration at the initial mass, so that the multiplier $\kappa$ is the only tuning parameter.
The terminal weight $\bs{P}_f$ is the stationary Riccati solution at the nominal arrival state, which keeps the gain continuous up to arrival and adds no parameter.
The sweep covered $\kappa\in\{0.7,1,1.5,2,3,5,10,20,50,100,200,500,1000,2000,5000\}$, bounded below by the feedback rate that can still be resolved at the control interval.
The selected multipliers are $\kappa=\num{500}$ for the HEO-to-$\mathcal{L}_1$-halo transfer and $\kappa=\num{2000}$ for the HEO-to-$\mathcal{L}_2$-halo transfer.

\appendixsection{Station-Keeping Verification of the Arrival Tolerance}
\label{sec: appendix station keeping}
To verify the arrival tolerance $\epsilon_a$, we test whether a state within it can be station-kept on the target orbit under the mission thrust bound.
For this, the LQR tracking law of Appendix~\ref{sec: appendix tracking baselines} tracks the target orbit for five periods at zero nominal thrust, with the same weights and control interval and the multiplier $\kappa=\num{1}$.
A state counts as recovered if it stays within $\epsilon_a$ of the orbit after five periods.
For selected values of $\epsilon_a$, we displace \num{2000} orbit states at uniformly random phases by $\epsilon_a$ in uniformly random six-dimensional directions.
For the HEO-to-$\mathcal{L}_1$-halo transfer, station keeping recovers every state at the chosen $\epsilon_a=\num{0.01}$ and \SI{98.25}{\percent} at $\epsilon_a=\num{0.02}$.
For the HEO-to-$\mathcal{L}_2$-halo transfer, it recovers \SI{99.85}{\percent} at the chosen $\epsilon_a=\num{0.02}$ and \SI{93.70}{\percent} at $\epsilon_a=\num{0.05}$.
The chosen tolerances therefore correspond to arrival states from which station keeping almost always holds the spacecraft on the orbit.
Without thrust, none of \num{20} states per orbit displaced by \num{0.01} ends within \num{0.01} of the orbit after five periods, even for the linearly stable $\mathcal{L}_1$ halo orbit, so the tolerances depend on active station keeping.
The fuel consumption for this station keeping depends on how close the spacecraft arrives and is not included in the added fuel of Tables~\ref{tab:l1_comparison} and~\ref{tab:l2_comparison}. 

\appendixsection{Onboard Computation}
\label{sec: appendix onboard computation}
All methods of this paper are implemented in Python and call the C++ core of \texttt{pydylan} for each propagation.
Their runtime on a workstation includes setting up every call and the Python code between the calls, not only the arithmetic, and does not transfer to the actual runtime on a flight computer.
We therefore count the FLOPs that a method executes per trajectory in a single-threaded run on an AMD EPYC 9654 processor for every test sample.
The count is the sum of the processor's hardware counters for SSE/AVX and x87 FLOPs, which we read through the Linux \texttt{perf\_event\_open} interface before and after each function.
It counts each element of a vector instruction as one FLOP and a fused multiply-add as two, covers the Python code, the numerical libraries, and \texttt{pydylan} alike, and excludes integer and memory operations.

The counts cover every replan of Algorithm~\ref{alg:costate-replanning} and leave out only one-time setup that can be done before flight, such as loading the network and propagating the target orbit.
As for the network, the PD and LQR counts include the simulated flight, and the feedback law itself takes only about \SI{25}{\percent} of their FLOPs.
The re-solve count includes all FLOPs of both solver runs on every sample, from SNOPT itself and from the propagations it requests.
For the Hard network rows of Tables~\ref{tab:l1_comparison} and~\ref{tab:l2_comparison}, inference takes about \SI{48}{\percent} of the FLOPs without refinement for the HEO-to-$\mathcal{L}_1$-halo transfer, but only \SI{10}{\percent} for the HEO-to-$\mathcal{L}_2$-halo transfer with its smaller network.
With refinement, the spacecraft arrives after fewer replans, so the network is queried less often, and with the added derivative evaluations, the share of inference falls to \SI{4}{\percent} and \SI{1}{\percent}.

To relate these counts to a flight computer, we divide them in Tables~\ref{tab:l1_comparison} and~\ref{tab:l2_comparison} by \SI[per-mode=symbol]{20}{\mega\flop\per\second}.
This is about the rate that one core of the radiation-hardened quad-core GR740 processor sustains on a floating-point benchmark, well below its peak of \SI[per-mode=symbol]{250}{\mega\flop\per\second}~\cite{Jalle.2016}.
The resulting times are estimates, because we assume this rate for every operation and omit memory effects.
An implementation optimized for this processor could lower the counts.
Newer space processors and dedicated accelerators for onboard neural-network inference, some of which have already flown, could shorten these times further~\cite{IzzoEtAl.2024}.
Storage does not limit the network size.
In single precision, the networks with $W=\num{200}$, $L=\num{5}$ and $W=\num{400}$, $L=\num{7}$ take \SI{0.66}{\mega\byte} and \SI{3.9}{\mega\byte}.
Size matters instead for computation, because one inference costs about two FLOPs per parameter, and for memory access, because only the smaller network fits in the \SI{2}{\mebi\byte} level-2 cache of the GR740.
A larger network is also harder to verify.

\begin{figure}
\centering
\includegraphics[width=0.85\linewidth]{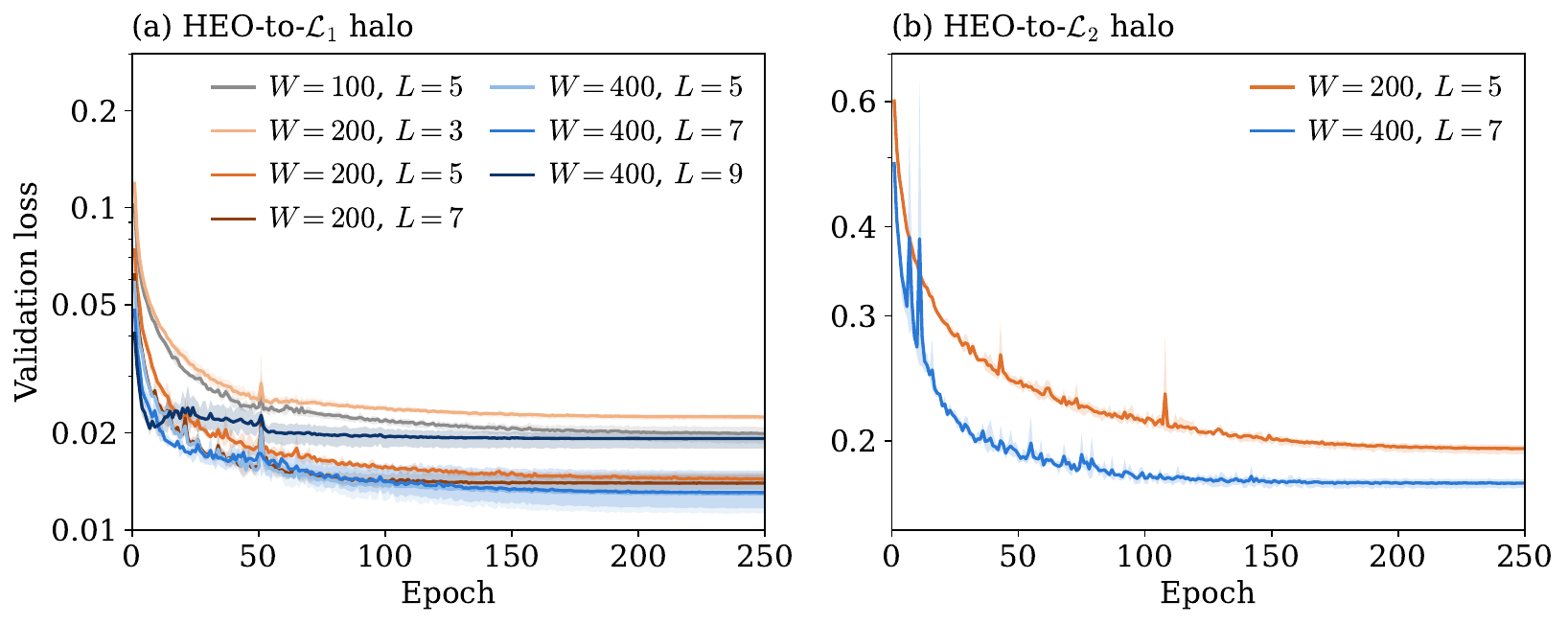}
\caption{Validation regression loss during training for the network sizes of the a) HEO-to-$\mathcal{L}_1$-halo and b) HEO-to-$\mathcal{L}_2$-halo transfers. Bands span three training seeds.}
\label{fig:validation-loss}
\end{figure}

\appendixsection{Design Studies}
\label{sec: appendix design studies}
Table~\ref{tab:l1_design_studies} lists all values of the HEO-to-$\mathcal{L}_1$-halo design studies of Sec.~\ref{sec: design studies}, including the Medium test case that Fig.~\ref{fig:design-studies} omits, and Table~\ref{tab:l2_design_studies} those of the smaller study with which we selected the dataset and network of the HEO-to-$\mathcal{L}_2$-halo transfer.
The Hard MFLOP column, defined in Sec.~\ref{sec: design studies}, is counted for a single training seed, whereas the arrival errors are averaged over the seeds.
Figure~\ref{fig:validation-loss} shows the validation regression loss of the size-study networks during training.
The spike at epoch \num{50} marks the start of the HEO-to-$\mathcal{L}_1$-halo Hamiltonian penalty.

\begin{table}
\caption{HEO-to-$\mathcal{L}_1$-halo design studies. Mean $d_f\pm$ one standard deviation over three training seeds. Boldface marks each block's lowest mean.}
\label{tab:l1_design_studies}
\centering
\setlength{\tabcolsep}{5pt}
\begin{tabular}{lS[table-format=7.0]S[table-format=1.4]S[table-format=1.4]S[table-format=1.4]S[table-format=3.1]}
\toprule
Setting & \multicolumn{1}{c}{Parameters} & \multicolumn{1}{c}{Easy} & \multicolumn{1}{c}{Medium} & \multicolumn{1}{c}{Hard} & \multicolumn{1}{c}{Hard [\si{\mega\flop}]} \\
\midrule
\multicolumn{6}{l}{\textit{Training dataset} ($M=\num{3000}$, $W=\num{200}$, $L=\num{5}$, $R=\num{10}$)} \\
D1 &  & \num{0.0234 \pm 0.0058} & \num{0.0312 \pm 0.0059} & \num{0.0366 \pm 0.0014} & \\
D2 &  & \num{0.0211 \pm 0.0015} & \num{0.0287 \pm 0.0022} & \num{0.0349 \pm 0.0022} & \\
D3 &  & \num{0.0228 \pm 0.0033} & \num{0.0313 \pm 0.0031} & \num{0.0344 \pm 0.0007} & \\
D4 &  & \num{0.0198 \pm 0.0020} & \num{0.0260 \pm 0.0032} & \num{0.0323 \pm 0.0016} & \\
D5 &  & \num{0.0186 \pm 0.0009} & \num{0.0254 \pm 0.0017} & \num{0.0294 \pm 0.0015} & \\
D6 &  & \bestnum{0.0142 \pm 0.0021} & \bestnum{0.0170 \pm 0.0025} & \bestnum{0.0232 \pm 0.0021} & \\
\midrule
\multicolumn{6}{l}{\textit{Anchors $M$} (D6, $W=\num{200}$, $L=\num{5}$, $R=\num{10}$)} \\
\num{750} &  & \num{0.0257 \pm 0.0029} & \num{0.0304 \pm 0.0041} & \num{0.0353 \pm 0.0027} & \\
\num{1500} &  & \num{0.0187 \pm 0.0024} & \num{0.0266 \pm 0.0023} & \num{0.0300 \pm 0.0014} & \\
\num{2250} &  & \num{0.0175 \pm 0.0041} & \num{0.0202 \pm 0.0038} & \num{0.0251 \pm 0.0005} & \\
\num{3000} &  & \num{0.0142 \pm 0.0021} & \bestnum{0.0170 \pm 0.0025} & \bestnum{0.0232 \pm 0.0021} & \\
\num{4500} &  & \bestnum{0.0130 \pm 0.0019} & \num{0.0226 \pm 0.0054} & \num{0.0279 \pm 0.0056} & \\
\midrule
\multicolumn{6}{l}{\textit{Network width $W$ and depth $L$} (D6, $R=\num{10}$)} \\
$W=\num{100}$, $L=\num{5}$ & 42007 & \num{0.0220 \pm 0.0038} & \num{0.0270 \pm 0.0020} & \num{0.0335 \pm 0.0009} & \\
$W=\num{200}$, $L=\num{3}$ & 83607 & \num{0.0275 \pm 0.0032} & \num{0.0330 \pm 0.0048} & \num{0.0371 \pm 0.0034} & \\
$W=\num{200}$, $L=\num{5}$ & 164007 & \num{0.0142 \pm 0.0021} & \num{0.0170 \pm 0.0025} & \num{0.0232 \pm 0.0021} & \\
$W=\num{200}$, $L=\num{7}$ & 244407 & \num{0.0111 \pm 0.0034} & \num{0.0199 \pm 0.0032} & \num{0.0240 \pm 0.0027} & \\
$W=\num{400}$, $L=\num{5}$ & 648007 & \num{0.0107 \pm 0.0011} & \num{0.0183 \pm 0.0015} & \num{0.0224 \pm 0.0012} & \\
$W=\num{400}$, $L=\num{7}$ & 968807 & \bestnum{0.0082 \pm 0.0006} & \bestnum{0.0139 \pm 0.0005} & \bestnum{0.0180 \pm 0.0008} & \\
$W=\num{400}$, $L=\num{9}$ & 1289607 & \num{0.0130 \pm 0.0063} & \num{0.0246 \pm 0.0163} & \num{0.0282 \pm 0.0144} & \\
\midrule
\multicolumn{6}{l}{\textit{Replans $R$} (D6, $W=\num{400}$, $L=\num{7}$, $n_r=\num{0}$)} \\
5 &  & \num{0.0152 \pm 0.0027} & \num{0.0235 \pm 0.0015} & \num{0.0270 \pm 0.0015} & 14.2 \\
10 &  & \num{0.0082 \pm 0.0006} & \num{0.0139 \pm 0.0005} & \num{0.0180 \pm 0.0008} & 22.2 \\
20 &  & \num{0.0065 \pm 0.0006} & \num{0.0095 \pm 0.0004} & \num{0.0141 \pm 0.0015} & 35.6 \\
30 &  & \bestnum{0.0062 \pm 0.0003} & \num{0.0083 \pm 0.0003} & \num{0.0125 \pm 0.0012} & 47.2 \\
50 &  & \num{0.0067 \pm 0.0005} & \bestnum{0.0081 \pm 0.0002} & \num{0.0115 \pm 0.0009} & 67.6 \\
100 &  & \num{0.0072 \pm 0.0002} & \num{0.0081 \pm 0.0004} & \bestnum{0.0110 \pm 0.0008} & 112.9 \\
\midrule
\multicolumn{6}{l}{\textit{Refinement iterations $n_r$} (D6, $W=\num{400}$, $L=\num{7}$, $R=\num{30}$)} \\
0 &  & \num{0.0062 \pm 0.0003} & \num{0.0083 \pm 0.0003} & \num{0.0125 \pm 0.0012} & 47.2 \\
1 &  & \num{0.0048 \pm 0.0003} & \num{0.0053 \pm 0.0004} & {\num[retain-zero-uncertainty]{0.0071 \pm 0.0000}} & 121.9 \\
2 &  & \bestnum{0.0046 \pm 0.0002} & \num{0.0051 \pm 0.0003} & \num{0.0062 \pm 0.0001} & 145.3 \\
3 &  & \num{0.0046 \pm 0.0003} & \bestnum{0.0050 \pm 0.0002} & \num{0.0058 \pm 0.0001} & 162.7 \\
4 &  & \num{0.0046 \pm 0.0002} & \num{0.0050 \pm 0.0002} & \bestnum{0.0056 \pm 0.0001} & 175.8 \\
\midrule
\multicolumn{6}{l}{\textit{Component removed} (D6, $W=\num{400}$, $L=\num{7}$, $R=\num{30}$, $n_r=\num{0}$)} \\
Full method &  & \num{0.0062 \pm 0.0003} & \bestnum{0.0083 \pm 0.0003} & \bestnum{0.0125 \pm 0.0012} & \\
No Hamiltonian penalty &  & \bestnum{0.0058 \pm 0.0003} & \num{0.0090 \pm 0.0015} & \num{0.0137 \pm 0.0023} & \\
No costate rescaling &  & \num{0.0117 \pm 0.0013} & \num{0.0169 \pm 0.0024} & \num{0.0233 \pm 0.0026} & \\
No anchors &  & \num{0.0095 \pm 0.0002} & \num{0.0818 \pm 0.0038} & \num{0.1040 \pm 0.0062} & \\
\bottomrule
\end{tabular}
\end{table}

\begin{table}
\caption{HEO-to-$\mathcal{L}_2$-halo selection study. Mean $d_f\pm$ one standard deviation over nine networks, three in the last two blocks. Boldface marks each block's lowest mean.}
\label{tab:l2_design_studies}
\centering
\begin{tabular}{lS[table-format=1.4]S[table-format=1.4]S[table-format=1.4]S[table-format=3.1]}
\toprule
Setting & \multicolumn{1}{c}{Easy} & \multicolumn{1}{c}{Medium} & \multicolumn{1}{c}{Hard} & \multicolumn{1}{c}{Hard [\si{\mega\flop}]} \\
\midrule
\multicolumn{5}{l}{\textit{Training dataset} ($W=\num{200}$, $L=\num{5}$, $R=\num{10}$)} \\
D1 & \num{0.1323 \pm 0.0377} & \num{0.1754 \pm 0.0181} & \num{0.2046 \pm 0.0157} & \\
D2 & \num{0.1146 \pm 0.0218} & \num{0.1750 \pm 0.0198} & \num{0.2047 \pm 0.0210} & \\
D3 & \bestnum{0.1134 \pm 0.0291} & \num{0.1688 \pm 0.0233} & \num{0.1995 \pm 0.0230} & \\
D4 & \num{0.1345 \pm 0.0329} & \num{0.1708 \pm 0.0248} & \bestnum{0.1963 \pm 0.0240} & \\
D5 & \num{0.1172 \pm 0.0255} & \bestnum{0.1645 \pm 0.0301} & \num{0.1970 \pm 0.0315} & \\
\midrule
\multicolumn{5}{l}{\textit{Anchors $M$} (D1 recipe, $W=\num{200}$, $L=\num{5}$, $R=\num{10}$)} \\
\num{750} & \bestnum{0.1299 \pm 0.0218} & \num{0.1915 \pm 0.0267} & \num{0.2175 \pm 0.0261} & \\
\num{1500} (D1) & \num{0.1323 \pm 0.0377} & \bestnum{0.1754 \pm 0.0181} & \bestnum{0.2046 \pm 0.0157} & \\
\num{2250} & \num{0.1314 \pm 0.0511} & \num{0.1874 \pm 0.0369} & \num{0.2132 \pm 0.0350} & \\
\midrule
\multicolumn{5}{l}{\textit{Network $W=\num{400}$, $L=\num{7}$} (D1 recipe, $R=\num{10}$)} \\
$M=\num{1500}$ & \bestnum{0.1489 \pm 0.0323} & \bestnum{0.1690 \pm 0.0210} & \bestnum{0.2008 \pm 0.0274} & \\
$M=\num{2250}$ & \num{0.1536 \pm 0.0200} & \num{0.1744 \pm 0.0278} & \num{0.2087 \pm 0.0249} & \\
\midrule
\multicolumn{5}{l}{\textit{Refinement iterations $n_r$} (D2, $W=\num{200}$, $L=\num{5}$, $R=\num{30}$)} \\
0 & \num{0.0620 \pm 0.0041} & \num{0.1299 \pm 0.0063} & \num{0.1604 \pm 0.0076} & 109.5 \\
1 & \num{0.0282 \pm 0.0063} & \num{0.0753 \pm 0.0085} & \num{0.0946 \pm 0.0075} & 407.1 \\
2 & \num{0.0246 \pm 0.0062} & \num{0.0574 \pm 0.0070} & \num{0.0775 \pm 0.0027} & 535.3 \\
3 & \num{0.0221 \pm 0.0053} & \num{0.0498 \pm 0.0039} & \num{0.0671 \pm 0.0015} & 618.4 \\
4 & \bestnum{0.0207 \pm 0.0041} & \bestnum{0.0451 \pm 0.0043} & \bestnum{0.0616 \pm 0.0029} & 679.1 \\
\midrule
\multicolumn{5}{l}{\textit{Component removed} (D2, $W=\num{200}$, $L=\num{5}$, $R=\num{30}$, $n_r=\num{0}$)} \\
Full method & \bestnum{0.0620 \pm 0.0041} & \num{0.1299 \pm 0.0063} & \num{0.1604 \pm 0.0076} & \\
No Hamiltonian penalty & \num{0.1085 \pm 0.0221} & \bestnum{0.1248 \pm 0.0313} & \bestnum{0.1537 \pm 0.0279} & \\
No costate rescaling & \num{0.2258 \pm 0.0340} & \num{0.1935 \pm 0.0384} & \num{0.2217 \pm 0.0284} & \\
No anchors & \num{0.1951 \pm 0.0070} & \num{0.3571 \pm 0.0302} & \num{0.3666 \pm 0.0292} & \\
\bottomrule
\end{tabular}
\end{table}

For the HEO-to-$\mathcal{L}_2$-halo transfer, each entry of the first three blocks of Table~\ref{tab:l2_design_studies} averages nine networks, three partitions of the anchors into training, validation, and test sets with three training seeds each, because the spread across partitions is as large as the differences between the variants.
None of the changes that reduce the error of the HEO-to-$\mathcal{L}_1$-halo transfer reduces it here by more than this spread.
Doubling the trajectories per anchor (D4), increasing the number of anchors, and enlarging the network to $W=\num{400}$ and $L=\num{7}$ all leave the Hard error within one standard deviation of the corresponding baseline.
Although the larger network fits the validation data better, as shown in Fig.~\ref{fig:validation-loss}b, this improvement in validation loss does not translate into better guidance performance.
Because no dataset improves on the others by more than this spread, we use D2, which has half the trajectories of D4 and D5.
Datasets D1 and D3 would be equally valid choices.

In the last two blocks, we repeat the refinement-iteration and component-removed studies of Table~\ref{tab:l1_design_studies}, averaged over three training seeds.
Refinement helps more than for the HEO-to-$\mathcal{L}_1$-halo transfer.
Four iterations lower the error by \SIrange{62}{67}{\percent} across the test cases, against \SIrange{26}{55}{\percent} there, and even the fourth iteration still lowers it by \SIrange{6}{9}{\percent}.
The network without anchors is trained on nominal-only data with the multipliers of D2.
As for the HEO-to-$\mathcal{L}_1$-halo transfer, removing the anchors raises the Medium and Hard errors the most, followed by removing the costate rescaling.
Removing the Hamiltonian penalty raises the Easy error by \SI{75}{\percent} but changes the Medium and Hard errors by less than the spread across training seeds, which is four to five times larger without the penalty.

\section*{Funding Sources}
The second author acknowledges partial support during the course of this effort from the Princeton Laboratory for Artificial Intelligence's [PLI/AI\textsuperscript{2}/NAM] initiative.

\section*{Acknowledgments}
Simulations were performed on computational resources managed and supported by Princeton Research Computing, a consortium of groups including the Princeton Institute for Computational Science and Engineering and the Office of Information Technology's High-Performance Computing Center and Visualization Laboratory at Princeton University.
Claude Code (Anthropic) and Codex (OpenAI) were used to perform final minor editing of language of the author-written manuscript and to assist with the plotting code.
The authors reviewed and verified all content and take full responsibility for this paper.

\bibliography{references}

\end{document}